\def\version{July 13, 2001}

\documentclass[reqno,11pt]{amsart}

\usepackage{amssymb, epsf}

\def\be{\begin{equation}}
\def\ba{\begin{align}}
\def\bm{\begin{multline}}
\def\bfig{\begin{figure}[htb]}
\def\efig{\end{figure}}

\newcommand{\bibit}[1]{\bibitem[#1]{#1}}
\newcommand{\paper}[1]{{\it #1}, }
\newcommand{\journal}[4]{#1 {\bf #2}, #3 (#4)}
\newcommand{\CMP}{Commun. Math. Phys.}
\newcommand{\HPA}{Helv. Phys. Acta}
\newcommand{\JSP}{J. Stat. Phys.}

\numberwithin{equation}{section}
\newtheorem{theorem}{Theorem}[section]

\newcommand{\fig}{Fig.\;}

\newcommand{\eg}{e.g.\;}
\newcommand{\ie}{i.e.\;}
\newcommand{\nn}{\nonumber}

\renewcommand{\leq}{\;\leqslant\;}
\renewcommand{\geq}{\;\geqslant\;}

\newcommand{\dd}{{\rm d}}
\newcommand{\e}[1]{\,{\rm e}^{#1}\,}
\newcommand{\ii}{{\rm i}}
\newcommand{\sumtwo}[2]{\sum_{\substack{#1 \\ #2}}}

\DeclareMathOperator*{\union}{\text{\large$\cup$}}

\newcommand{\fall}{\;\forall\,}

\def\Tr{{\operatorname{Tr\,}}}

\def\dist{{\operatorname{dist\,}}}

\def\low{{\operatorname{low\,}}}
\def\high{{\operatorname{high\,}}}
\def\Re{{\operatorname{Re\,}}}

\def\per{{\text{\rm\,per}}}

\def\bra #1{\langle#1 |\,}
\def\ket #1{\,|#1 \rangle}

\newcommand{\expval}[1]{\langle #1 \rangle}

\newcommand{\neighbours}[2]{<\! #1,#2 \! >}
\newcommand{\compl}{{\text{\rm c}}}

\newcommand{\const}{{\text{\rm const}}}

\def\writefig#1 #2 #3 {\rlap{\kern #1 truecm \raise #2 truecm
\hbox{#3}}}
\def\figtext#1{\smash{\hbox{#1}} \vspace{-5mm}}
\newcommand{\caA}{{\mathcal A}}
\newcommand{\caB}{{\mathcal B}}
\newcommand{\caC}{{\mathcal C}}

\newcommand{\caE}{{\mathcal E}}
\newcommand{\caF}{{\mathcal F}}
\newcommand{\caG}{{\mathcal G}}
\newcommand{\caH}{{\mathcal H}}

\newcommand{\caL}{{\mathcal L}}

\newcommand{\caN}{{\mathcal N}}
\newcommand{\caO}{{\mathcal O}}
\newcommand{\caP}{{\mathcal P}}
\newcommand{\caQ}{{\mathcal Q}}

\newcommand{\caU}{{\mathcal U}}

\newcommand{\caW}{{\mathcal W}}

\newcommand{\bbC}{{\mathbb C}}

\newcommand{\bbR}{{\mathbb R}}

\newcommand{\bbZ}{{\mathbb Z}}

\newcommand{\frG}{{\mathfrak G}}

\newcommand{\frM}{{\mathfrak M}}

\newcommand{\bsB}{{\boldsymbol B}}

\newcommand{\bsE}{{\boldsymbol E}}

\newcommand{\bsH}{{\boldsymbol H}}

\newcommand{\bsK}{{\boldsymbol K}}
\newcommand{\bsL}{{\boldsymbol L}}

\newcommand{\bsR}{{\boldsymbol R}}
\newcommand{\bsS}{{\boldsymbol S}}
\newcommand{\bsT}{{\boldsymbol T}}

\newcommand{\bsV}{{\boldsymbol V}}

\newcommand{\bsmu}{{\boldsymbol \mu}}
\newcommand{\bsomega}{{\boldsymbol \omega}}
\begin{document}

{\small
%\framebox{PREPRINT}
\hfill \version
}
\vspace{2mm}

\title{Quantum lattice models at intermediate temperature}
\maketitle

\renewcommand{\thefootnote}{\roman{footnote}}

\begin{centerline}{\sc
J. Fr\"ohlich$^1$\footnote{Institut f\"ur Theoretische Physik, ETH
H\"onggerberg, CH--8093 Z\"urich; juerg@itp.phys.ethz.ch.},
L. Rey-Bellet$^2$\footnote{Dept of Mathematics, University of
Virginia, Charlottesville, VA 22903; lr7q@virginia.edu},
D. Ueltschi$^3$\footnote{Dept of Physics, Princeton University, Jadwin
Hall, Princeton, NJ 08544; ueltschi@princeton.edu. D.U.\ is partially
supported by the US National Science Foundation, grant PHY 9820650.}
}
\end{centerline}

\setcounter{footnote}{0}
\renewcommand{\thefootnote}{\arabic{footnote}}

\vspace{5mm}

\begin{centering}
{\small\it
$^1$ETH, Z\"urich, Switzerland\\
$^2$University of Virginia, USA\\
$^3$Princeton University, New Jersey, USA\\
}
\end{centering}

\vspace{5mm}

\begin{quote}
{\small {\bf Abstract.} We analyze the free energy and construct the 
Gibbs-KMS states for a class of quantum lattice systems, at low 
temperature and when the interactions are almost diagonal, in a suitable 
basis. The models we study may have continuous symmetries, our results 
however apply to intermediate temperatures where discrete symmetries are 
broken but continuous symmetries are not. Our results are based on 
quantum Pirogov-Sinai theory and a combination of high and low temperature 
expansions.}

\vspace{1mm}

\vspace{3mm}

{\it Dedicated to Joel Lebowitz on the occasion of his seventieth birthday.}

\end{quote}

\vspace{3mm}

\section{Introduction} In this paper we study the low temperature phase 
diagram for a class of quantum lattice systems. Starting with
\cite{PS, Sin}, Pirogov-Sinai theory has evolved \cite{KP, Zah, BKL,
BS, BI, BK} into a very powerful tool to study the pure phases, their
coexistence and the first-order phase transitions in {\em classical}
spin systems at low temperature.  In recent years large part of the
Pirogov-Sinai theory has been extended to quantum systems \cite{Pir,
BKU, DFF, DFFR, KU}, quantum spin systems as well as
fermionic and bosonic lattice gases, and applied to a variety of models
\cite{FR, DFF2, GKU} to describe insulating phases associated with
discrete symmetry breaking. Here we formulate the Pirogov-Sinai theory
in terms of tangent functionals to the free energy. This allows to
discuss the completeness of the phase diagram avoiding the
difficulties associated with boundary conditions.  We reformulate
results of \cite{BKU, DFF, DFFR, KU} in this framework, and extend the
theory to a class of models where discrete symmetries are broken at
intermediate temperatures. This applies in particular to some systems
with continuous symmetries. For this, we consider the {\em restricted
ensembles} introduced in \cite{BKL} that are very useful to analyze
phases which are associated to a family of configurations rather than
to a single configuration.

The models that we consider have Hamiltonians, for finite volumes
$\Lambda$, of the form
\begin{displaymath}
H_\Lambda \,=\, V_\Lambda + T_\Lambda
\end{displaymath}
where $V_\Lambda$ is a classical Hamiltonian (i.e. diagonal in a
suitable basis) and $T_\Lambda$ is a (usually small) quantum
perturbation. In typical situations the suitable basis is the basis of
occupation numbers of position operators. Electronic systems provide a
large class of interesting models. The classical interaction
$V_\Lambda$ describes the many-body short range and classical
interaction between the spin-$\frac12$ fermions as well as external
fields and chemical potentials:
\begin{displaymath}
V_\Lambda \,=\,\ \sumtwo{x \in \Lambda}{\sigma \in \{\uparrow,\downarrow\}} 
J_{x,\sigma} n_{x,\sigma} + \sumtwo{x,y \in \Lambda}{\sigma, \sigma' \in\{ 
\uparrow,\downarrow\}} J_{xy,\sigma\sigma'} n_{x,\sigma} n_{y,\sigma'} + 
\cdots \,.
\end{displaymath}
A typical quantum perturbation $T_\Lambda$ is the kinetic energy
\begin{displaymath}
T_\Lambda\,=\,  \sumtwo{\neighbours xy \subset \Lambda}
{\sigma \in \{\uparrow,\downarrow\}} t_{xy,\sigma} (c^\dagger_{x\sigma} 
c_{y\sigma} + {\rm h.c.})\,,
\end{displaymath}
where $c^\dagger_{x\sigma}$ and $c_{x\sigma}$ are the creation and
annihilation operators and $\neighbours xy$ denotes pairs of nearest 
neighbors.

Often, in such systems, the behavior at low temperatures arises from a 
subtle interplay between the (classical) potential energy and the kinetic 
energy. In this paper two  such mechanisms are considered and combined, 
each of which we now illustrate with an example. 

\medskip
\noindent
{\bf Example 1: Hubbard Model}. In this case the (classical) interaction is 
only on-site: 
\begin{displaymath} 
V_\Lambda \,=\, \sum_{x \in \Lambda} U n_{x \uparrow}  n_{x \downarrow} - \mu 
(n_{x \uparrow} +  n_{x \downarrow})\,.  
\end{displaymath}
For suitable values of $U$ and $\mu$, the ground states of $V_\Lambda$
have an infinite degeneracy (in the thermodynamic limit): each site is
occupied by a single particle of arbitrary spin.  However the kinetic
energy lifts this degeneracy and induces an effective
antiferromagnetic interaction between nearest neighbors.  The
perturbative methods of \cite{DFFR,DFF2} shows that, in this parameter
range, this system is equivalent, in the sense of statistical
mechanics, to the Heisenberg antiferromagnet, up to controlled error
terms.  If the hopping coefficients are asymmetric (e.g. $t_{xy,
\uparrow} \ll t_{xy, \downarrow}$) then quantum Pirogov-Sinai implies
the coexistence of two antiferromagnetic phases at low enough
temperatures \cite{DFFR, KU, DFF2}. Rigorous results for the Hubbard
model are reviewed in \cite{Lieb}.

\medskip
\noindent {\bf Example 2: Extended Hubbard Model}. This variant of the 
Hubbard model includes a nearest neighbor interaction: 
\begin{displaymath}
V_\Lambda \,=\, \sum_{x \in \Lambda} \Bigl[ U n_{x \uparrow}  n_{x \downarrow} - \mu 
(n_{x \uparrow} +  n_{x \downarrow}) \Bigr] + W \sum_{\neighbours xy \subset \Lambda}
(n_{x \uparrow} +  n_{x \downarrow})  (n_{y \uparrow} +  n_{y \downarrow})
\,. 
\end{displaymath}
If the the interaction between nearest neighbors is repulsive then for
suitable values of $U$, $W$ and $\mu$ the ground states of $V_\Lambda$ are
chessboard configurations where empty sites alternate with sites
occupied with one particle of arbitrary spin. The degeneracy of the ground states is
infinite in the thermodynamic limit but we have a spatial ordering of the 
particle. Using restricted ensemble we associate a pure phase to this 
spatial ordering by neglecting the spin degrees of freedom. The methods of 
this paper imply the existence of only two pure phases in the intermediate 
temperature range  
\begin{displaymath}
\beta t \ll 1 \quad {\rm ~and~} \quad \beta W \gg 1 \,.
\end{displaymath}
The temperature is so low that the spatial ordering of
the particles survives but so high that the spin are
in a {\em disordered} phase. The continuous symmetry (if $t_{xy,
\uparrow} = t_{xy, \downarrow}$) is {\em not} broken in this parameter
regime.

These two models illustrate some of the mechanisms arising from the 
competition between classical and quantum effects, where the system 
remains insulating and no continuous symmetry is broken.  
Our main result, Theorem \ref{thmLSintemp}, provides tools to describe
the phase diagram of such models, in particular the 
coexistence of several phases and the associated first-order phase 
transitions. 

The main technical ingredient in this paper is a combined low-temperature and 
high-temperature expansion for suitable contour models obtained using the 
perturbation theory developed in \cite{DFFR}. 

This paper is organized as follows. In Section \ref{secgenfr} we
describe the general formalism of quantum lattice systems and the
perturbation theory of
\cite{DFFR}. Section \ref{secPS} is devoted to the Pirogov-Sinai theory. 
In Section \ref{secresults} we state the results of 
Pirogov-Sinai theory for quantum systems. The extended Hubbard model
is discussed in Section \ref{secexample} as an illustration. In Section 
\ref{secexp} we prove our main result by studying a
contour model and deriving the required bounds on the contours.

\section{General framework of quantum lattice models}
\label{secgenfr}

\subsection{Basic set-up} 
\label{sset} 

We consider a quantum mechanical system on a $\nu$-dimensional lattice
$\bbZ^{\nu}$, as considered, e.g., in \cite{Rue,Isr,BR,Sim}.  We will
need a slight modification of the usual formalism in order to treat
fermionic lattice gases \cite{DFFR} and to accommodate
the fact that fermionic creation and annihilation operators do not
commute but anticommute.  A quantum lattice system is defined by the
following data:
\medskip

\noindent
(i) {\bf Hilbert space.} For convenience we choose a total ordering
(denoted by the symbol $\preceq$) of the sites in $\bbZ^{\nu}$.  We
choose the {\em spiral order}\/, depicted in Figure \ref{order} for
$\nu=2$, and an analogous ordering for ${\nu} \ge 3$. This ordering
has the property that, for any finite set $A$, the set $\overline{A}
:= \{z \in \bbZ^{\nu}, z \preceq A \}$ of lattice sites which are
smaller than $A$, or belong to $A$, is finite.  To each lattice site
$a\in\,\bbZ^{\nu}$ is associated a finite-dimensional Hilbert space
$\caH_a$ and, for any finite subset $A=\{a_1 \prec \cdots \prec a_n\}
\subset \bbZ^{\nu}$, the corresponding Hilbert space $\caH_A$  is 
given by the ordered tensor product
\begin{equation} 
\caH_A\,=\, \caH_{a_1} \otimes \cdots \otimes \caH_{a_n} \;. 
\label{basis}
\end{equation}
We further require that there be a Hilbert space isomorphism 
$\phi_a :{\caH}_a \longrightarrow {\caH}$, for all $a \in \bbZ^{\nu}$.  
\begin{figure}
%\vspace{4cm}
\begin{center}
\begin{picture}(100,75)(-50,-25)
\put(0,0){\circle*{5}}
\put(0,8){\makebox(0,0){$1$}}%{$x_1\!\!=\!\!0$}}%\overline x$}}
\put(0,-8){\makebox(0,0){$(0\,,0)$}}
\put(25,0){\circle*{5}}
\put(25,8){\makebox(0,0){$2$}}
\put(25,25){\circle*{5}}
\put(25,33){\makebox(0,0){$3$}}
\put(0,25){\circle*{5}}
\put(0,33){\makebox(0,0){$4$}}
\put(-25,25){\circle*{5}}
\put(-25,33){\makebox(0,0){$5$}}
\put(-25,0){\circle*{5}}
\put(-25,8){\makebox(0,0){$6$}}
\put(-25,-25){\circle*{5}}
\put(-25,-17){\makebox(0,0){$7$}}
\thicklines
\put(0,-25){\vector(1,0){50}}
\put(50,-25){\vector(0,1){75}}
\put(50,50){\vector(-1,0){100}}
\put(-50,50){\vector(0,-1){37.5}}
\end{picture}
\end{center}
\caption{Spiral order in $\bbZ^2$}
\label{order}
\end{figure}
\medskip

\noindent
(ii) {\bf Field and observable algebras.} For any finite subset $A
\subset \bbZ^{\nu}$ an operator algebra ${\caF}_A$, the {\em field
algebra}, is given.  The algebra ${\caF}_A$ is isomorphic to the algebra
$\caB(\caH_A)$ of bounded operators on $\caH_A$, but in general
${\caF}_A \not =\caB (\caH_A)$, rather $\caF_A \subset
\caB(\caH_{\overline A})$.  The algebra ${\caF}_A$ is a
$*$-algebra equipped with a $C^*$-norm obtained from the operator norm
on $\caB(\caH_A)$.  If $A \subset B$ and $a \prec b$, for all $a \in
A$ and all $b \in B \setminus A$, then there is a natural embedding of
${\caF}_A$ into ${\caF}_B$: An operator $K \in {\caF}_A$ corresponds
to the operator $K \otimes {\bf 1}_{{\caH }_{B\backslash A}}$ in
${\caF}_B$. In the following we denote by $K$ both operators.  For the
infinite system the field algebra is the $C^*$ algebra given by
\begin{equation}
{\caF} \,=\, \overline{ \bigcup_{A \nearrow \bbZ^{\nu}}
{\caF}_A}^{\,{\rm norm}}\; ,
\end{equation}
(the limit being taken through a sequence of increasing subsets of
$\bbZ^{\nu}$, where increasing refers to the (spiral) ordering defined
above). 

\noindent
The algebras $\caF_A$ contain the {\em observable algebras} $\caO_A$ which
have the same embedding properties as the field algebras and, moreover,
satisfy the following commutativity condition: If $A \cap B =
\emptyset$, then for any $K \in {\caF}_A$, $L \in {\caO}_B$ we have
\begin{equation}
[K,L]\,=\,0\;. 
\label{coco}
\end{equation}
For the infinite system the observable algebra $\caO$ is given by 
\begin{equation}
{\caO} \,=\, \overline{ \bigcup_{A \nearrow \bbZ^{\nu}}
{\caO}_A}^{\,{\rm norm}}\; .
\end{equation}
The group of space translations $\bbZ^{\nu}$ acts as a
$*$-automorphism group $\{\tau_a \}_{a \in \bbZ^\nu}$ on the
algebras ${\caF}$ and ${\caO}$, with
\begin{equation} 
{\caF}_{X+a} = \tau_{a}({\caF}_{X}), \;\;\;\; {\caO}_{X+a} =
\tau_{a}({\caO}_{X}),  
\end{equation} 
for any $X\subset{\bbZ^{\nu}}$ and $a\in {\bbZ^{\nu}}$.
\medskip

\noindent
(iii) {\bf Interactions, dynamics and free energy} An interaction
$\bsH = \{ H_A\}$ is given: This is a map from the finite sets
${A\subset \bbZ^{\nu}}$ to self-adjoint operators $H_A$ in the {\em
observable algebra} ${\caO}_A$.  We assume the interaction to be
translation invariant or periodic, i.e., there is a lattice $\Gamma
\subseteq \bbZ^\nu$, with ${\rm dim} \Gamma = \nu$, such that $\tau_a
H_A = H_{A+a}$, for all $a \in \Gamma$ and all $A \subset \bbZ^\nu$.  
We will consider finite range or exponentially decaying interactions.
The norm of an interaction is defined as
\begin{equation}
  \label{s.int}
\|\bsH\|_r \,=\, \sup_{a \in \bbZ^\nu}
\sum_{A\ni a} \|H_A\| \e{r|A|}\,, 
\end{equation}
for some $r > 0$. Here $|A|$ denotes the cardinality of the
smallest connected subset of $\bbZ^{\nu}$ which contains $A$.  We
shall denote by $ {\caB}_r = \{ \bsH \colon \|\bsH \|_r <
\infty \}$ the corresponding Banach space of interactions. 

For a finite box $\Lambda$, we denote $H_\Lambda$ the finite-volume
Hamiltonian given by $H_\Lambda = \sum_{A\subset \Lambda} H_A$. Here,
we consider only periodic boundary conditions, \ie $\Lambda$ is the
$\nu$-dimensional torus $(\bbZ / L\bbZ)^\nu$, $L$ being the size of
$\Lambda$. In the sequel we will consider infinite volume limits; the notation
$\lim_{\Lambda \nearrow \bbZ^\nu}$ will stand for $\lim_{L\to\infty}$.

If $\bsH \in {\caB}_r$, the interaction $\bsH$ determines a one-parameter 
group of $*$-automorphisms, $ \{ \alpha_t\}_{t\in\bbR}$ on $\caF$. These 
automorphisms are constructed as the limit (in the strong topology) 
of the automorphisms $\alpha_t^\Lambda$ given by for 
$K \in \caF_A$, $A\subset\Lambda$ by
\be
\alpha_t^\Lambda(K)\,=\, \e{\ii tH_\Lambda} K \e{-\ii tH_\Lambda}\,.
\end{equation}
The proof is standard (see e.g. \cite{BR}). Note that one makes
crucial use of the commutativity condition \eqref{coco}.

For an interaction $\bsH$ and at inverse temperature $\beta$ the
partition function is defined as
\be
Z_\Lambda^\beta = \Tr \e{-\beta H_\Lambda};
\label{deffpart}
\end{equation}
the {\it free energy} $f(\bsH)$ is then
\begin{equation}
f(\bsH)\,=\,  -\frac1\beta \lim_{\Lambda \nearrow \bbZ^\nu} 
\frac1{|\Lambda|} \log Z^\beta_\Lambda.
\end{equation} 
Existence of the limit is a well-known result, see \cite{Isr,Sim}.
Notice that $f(\bsH)$ is a concave function of the interaction $\bsH$.
\medskip

\noindent
(iv) {\bf KMS states and tangent functionals.} 
A {\em state} $w$ on $\caO$ is a positive normalized linear functional 
on $\caO$. A state $w$ is {\em periodic} if 
$w \circ  \tau_a = w$, for all $a$ in a lattice 
$\Gamma\subset \bbZ^\nu$ and invariant if $\Gamma=\bbZ^\nu$. 
A KMS state at inverse temperature $\beta$ is a state 
$w_\beta$ which satisfy the KMS 
condition
\begin{equation}
w_\beta ( K \alpha_t (L) ) \,=\, 
w_\beta ( \alpha_{t-\ii\beta} (K) L )\,.
\end{equation} 
For finite systems with periodic boundary conditions it is easy to
check that the Gibbs state given by 
\be
w_{\beta\Lambda}(\,{\cdot}\,) \,=\, ({\rm Tr}\e{-\beta H_{\Lambda}})^{-1} 
{\rm Tr}(\e{-\beta H_{\Lambda}} \,{\cdot}\,)
\end{equation}
satisfies the KMS condition.  The set of KMS
states is convex, and $w$ is called {\em extremal} if it cannot
be written as linear combination of KMS states.  The state $w$ is
{\em clustering} if
\be
\lim_{ a \rightarrow \infty} w ( K \tau_a (L) ) \,=\, 
w(K) w(\tau_a L)\,,
\end{equation}
for all $K$, $L \in \caO$.
Note that a state $w$ is extremal if it is clustering. The state
$w$ is {\em exponentially clustering} if, for any local observables
$K\in \caO_A $, $L \in \caO_B$ we have the property
\be
w ( K \tau_a (L) ) - w(K) w(\tau_a L) \, \leq \, C_{K,L}
\e{- |a| / \xi}
\end{equation}
with $\xi>0$; here $C_{K,L}$ depends on $K$ and $L$ only.

If we consider the free energy as a function of the interaction, KMS
states at inverse temperature $\beta$ are in one-to-one correspondence
with tangent functionals to the free energy. The free energy $f$ is a
concave function of the interaction $\bsH$ and a linear functional $\alpha$
on $\caB_r$ is said to be tangent to $f$ at $\bsH$ if for all
interaction $\bsK \in \caB_r$ we have
\begin{equation}
f(\bsH + \bsK) \, \leq \, f(\bsH) + \alpha(\bsK)\,.
\end{equation}
To an invariant state $w$ we associate a tangent functional $\alpha$ 
defined by 
\begin{equation}
\alpha(\bsK) \,=\,  w( A_\bsK ) 
\label{corr}
\end{equation}
where $A_\bsK = \sum_{X \ni 0} |X|^{-1} K_X$ (and similarly for
periodic states).  The results of Israel and Araki
\cite{Isr,Ara} show that if $\alpha$ is a tangent functional at $\bsH$, 
then the invariant state $w$ defined in \eqref{corr} is a KMS
state at temperature $\beta$ and, conversely, for any KMS state at
temperature $\beta$ there is a unique tangent functional $\alpha$.
The identification of KMS states with tangent functionals will be very
useful to describe the phase diagrams arising from Pirogov-Sinai
theory.
\medskip

\noindent
{\bf Example.} As an illustration of the general formalism we consider
spin $1/2$ fermions, as in the examples treated in this paper. The
Hilbert space $\caH_a$ is isomorphic to $\bbC^4$. We denote
$c^\dagger_{a\sigma}$ and $c_{a\sigma}$ the creation and annihilation
operators of a particle at site $a$ with spin $\sigma \in
\{\uparrow, \downarrow\}$. One can construct an explicit representation 
of the creation and annihilation operators as operators in $\caB
(\caH_{ \overline{a}})$, see \eg Section 4.2 in \cite{DFFR}, but
$c_{a\sigma}, c^\dagger_{a\sigma} \notin \caB(\caH_a)$. The algebras $\caF_A
\subset \caB(\caH_{\overline{A}})$ are chosen to be the algebras generated by 
$c_{a\sigma}$, $c^\dagger_{a\sigma}$, $a\in A$, $\sigma \in
\{\uparrow, \downarrow\}$. The observable algebra $\caO_A$ are chosen as 
the algebras generated by {\em pairs} of creation or annihilation
operators.  It is easy to check that the elements 
$\caF_A$ and $\caO_A$ satisfy the commutativity condition
\eqref{coco}.
\medskip

\noindent
{\bf Classical interactions}
A particular class of interactions consists of the {\em classical
interactions}.  Let $\{e_{j}\}_{j \in I}$ be an
orthonormal basis of ${\caH}$. Then, for $A \subset \bbZ^{\nu}$,
\begin{equation}
{\caE}_{A} \,=\, \{ \otimes_{a \in A} e_{j_a}^{a} \}\,,\;\;\;{\rm with} 
\;\;e_{j_a}^{a} = \phi_{a}^{-1}\,e_{j}\;,
\end{equation}
is an orthonormal basis of ${\caH}_A$. We denote by ${\caC}({\caE}_{A})$
the abelian subalgebra of ${\caO}_A$ consisting of all operators 
which are diagonal in the basis ${\caE}_{A}$.
An interaction $\bsV$ is called {\it classical}, if there exists a
basis $\{e_{j}\}_{j \in I}$ of ${\caH}$ such that
\begin{equation}
V_A \in {\caC}({\caE}_{A}), \;\;\; {\rm for}\;\;{\rm all}\;\; A \subset 
\bbZ^{\nu}\;.
\end {equation}
The set $\Omega_{A}$ of configurations in $A$ is defined as
the set of all assignments $\{j_a\}_{\{a \in A\}}$ of an element $j_a
\in I$ to each $a$.  A configuration $\omega_{A}$ is an element
in $\Omega_{A}$.  There is a one-to-one correspondence between basis
vectors $ \bigotimes_{a \in A} e_{j_a}^{a}$ of ${\caH}_A$ and 
configurations on $A$: 
\begin{equation} 
\bigotimes_{a \in A} e_{j_a}^{a} \longleftrightarrow \omega_{A} \equiv 
\{j_a \}_{a \in A}\;. 
\label{rrr.1} 
\end{equation} 
In the sequel we shall use the notation $e_{\omega_A}$ to denote the
basis vector defined by the configuration $\omega_A$ via the
correspondence \eqref{rrr.1}.
Since a classical interaction $\bsV$ only depends on the numbers
\begin{equation}
\Phi_A(\omega_A) \,=\, \langle e_{\omega_A} | V_A | e_{\omega_A} \rangle
\end{equation}
we may view $\Phi_A$ as a (real-valued) function on the set of
configurations.  
Similarly the algebra ${\caC}({\caE}_A)$ may be viewed as the $*$-algebra of 
complex-valued functions on the set of configurations $\Omega_A$.

\subsection{Perturbation theory for interactions}
\label{secperturb}

The interactions we will study have the form $\bsH=\bsV +
\lambda\bsT$ where $\bsV$ is a classical interaction, $\bsT$ is a
perturbation and $\lambda$ a small parameter. A typical situation is
the following: the classical part of the interaction has infinitely
many ground states, i.e the number of ground states of the finite-volume
Hamiltonian $H_\Lambda$ diverges as $|\Lambda| \rightarrow \infty$,
but the perturbation $V$ lifts this degeneracy (completely or
partially). This is usually easy to check this using standard
perturbation theory for the finite-volume Hamiltonian
$V_\Lambda+\lambda T_\Lambda$. Standard perturbation theory however
does not work in the thermodynamic limit, the norm of the error
growing with $|\Lambda|$ and other methods are required.  Such methods
have been developed in \cite{DFFR} and applied in \cite{FR,DFF2} (see
also \cite{KU} for an alternative approach).

The idea is to construct an interaction ${\tilde \bsH}$ which is
equivalent to $\bsH$ and which can be cast in the form
\begin{equation}
{\tilde \bsH}\,=\, {\tilde \bsV(\lambda)} + {\tilde \bsT}(\lambda)\,,
\end{equation}
where now the degeneracy of the ground states of ${\tilde \bsV}$ is lifted and 
${\tilde \bsT}(\lambda)$ is suitably small with respect to 
${\tilde \bsV}(\lambda)$. 

Recall that two interactions 
$\bsH$ and ${\tilde \bsH}$ are {\em equivalent} if there exists a 
$*$-automorphism of the algebra $\caO$ of local observables such that 
\begin{equation}
\tilde H_A\,=\, \gamma(H_A)\,,
\end{equation}
for all $A$. In particular, if $\bsH \in \caB_r$, 
there exists ${\tilde r}$ such that ${\tilde \bsH} \in \caB_{\tilde r}$. 
A convenient way of constructing equivalent interactions is with a family 
of unitary transformations $U_\Lambda$. Let $S_A$, $A \subset \bbZ^\nu$, 
be a family of {\em antiselfadjoint} operators, 
periodic or translation invariant, with $S_A \in \caO_A$ and 
$\|\bsS\|_r < \infty$ for some $r >0$. We set 
$S_\Lambda = \sum_{A \subset \Lambda} S_A$ and then 
$U_\Lambda = \exp (S_\Lambda)$ is unitary. It is shown in \cite{DFFR} 
that if $\|\bsS\|_r$ is small enough then the unitary equivalent Hamiltonians 
${\tilde H}_\Lambda=U_\Lambda H U^{-1}_\Lambda$ define an interaction 
${\tilde \bsH} \in \caB_{{\tilde r}}$ for some ${\tilde r}>0$ and 
${\tilde \bsH}$ is equivalent to $\bsH$.

We consider now an interaction of the form $\bsH=\bsV+\lambda \bsT$
which satisfy the following conditions

\begin{itemize}
\item[{\bf (P1)}] 
The interaction $\bsV$ is classical and of finite range.  
Moreover, we assume that $V$ is given by a 
translation-invariant $m$-potential.
This last condition means that we can assume (if necessary by passing to a 
physically equivalent interaction) that there exists at least one 
configuration $\omega$ minimizing all $\Phi_{0X}$, i.e., 
\begin{equation} 
\Phi_{0X}(\omega) = 
\min_{\omega'}\,\Phi_{0X}(\omega') \;,  
\label{mpot} 
\end{equation} 
for all $ X $. 
For any {\it{m}}-potential, the set of all configurations for which 
Eq.\ \eqref{mpot} holds is the set of ground states of $\Phi_0$.   
\item[{\bf (P2)}] 
The perturbation interaction $\bsT$ is in some space Banach space 
${\caB}_r$ for some $r>0$. 
\end{itemize}
Since, by condition {\bf (P1)}, the ground states can be determined locally, 
there is a corresponding decomposition of the Hilbert space $\caH_A$ for all 
$A$: 
\be
\caH_A\,=\, \caH_A^\low \oplus \caH_A^\high\,,
\label{decomp}
\end{equation}
where $\caH_A^\low$ is the subspace spanned by the ground states of
$\bsV$. We can decompose any operator $K_A \in \caB(\caH_A)$ according to
their action on $\caH_A^\low$ and  $\caH_A^\high$: 
\be
K_A\,=\, K_A^{\rm ll} + K_A^{\rm hh} + K_A^{\rm lh} \,,  
\end{equation} 
with
\begin{eqnarray}
K_A^{\rm ll} \caH_A^\low \subset \caH_A^\low  &&
K_A^{\rm ll} \caH_A^\high = 0 \,, \nn \\
K_A^{\rm hh} \caH_A^\high \subset \caH_A^\high &&
K_A^{\rm hh} \caH_A^\low = 0 \,, \nn \\
K_A^{\rm lh} \caH_A^\low \subset \caH_A^\high &&
K_A^{\rm lh} \caH_A^\high \subset \caH_A^\low \,.
\label{structLS}
\end{eqnarray} 
Accordingly we decompose any interaction $\bsT$:
\be
\bsT\,=\, \bsT^{\rm ll} + \bsT^{\rm hh} + \bsT^{\rm lh} 
\end{equation}  

The following theorem shows that, for any integer $n \ge 1$, it is 
possible to construct an interaction $\bsH^{(n)}$ equivalent to $\bsH$ with 
the property that $\bsH^{(n)}$ is block diagonal up to order $n$. Note that 
this is a constructive result and a algorithm is given in \cite{DFFR} which 
allows to construct the unitary transformations $U^{(n)}_\Lambda$ and the 
interactions $\bsH^{(n)}$. 

\begin{theorem} 
\label{hh.th}
Consider an interaction of the form
\begin{equation}
\bsH\;=\;\bsV + \lambda \bsT
\end{equation}
where $\bsV$ satisfies Condition {\bf (P1)} and $\bsT$ satisfies
Condition {\bf (P2)}. For any integer $n\ge1$ there is $r_n>0$ and
$\lambda_n>0$ such that for $|\lambda| < \lambda_n$ there is an
interaction $\bsH^{(n)}= \bsV + \bsT^{(n)} \in \caB_{r_n}$, equivalent
to $\bsH$, with 
\be
\|\bsT^{(n)\rm lh} \|_{r_n} \,=\, {\rm O}(\lambda^{n+1})\,. 
\end{equation}
\end{theorem}

This theorem is useful to
analyze the low temperature behavior of quantum spin systems when the
ground states of $\bsV$ have infinite degeneracy and $\bsT$ lifts this
degeneracy (totally or partially). Consider for example the typical case 
where the degeneracy is lifted in second order perturbation 
theory. In that case we may take $n=1$ and we have 
$\bsT^{(1)\rm lh}=O(\lambda^2)$: 
\be
\bsH^{(1)} \,=\, \bsV + \sum_{j \ge 1} \lambda^j \bsT^{(1)\rm ll}_j +
\sum_{j \ge 1}\lambda^j \bsT^{(1)\rm hh}_j + \sum_{j \ge 2} \lambda^j 
\bsT^{(1)\rm lh}_j \,.
\end{equation}
We then decompose $\bsH^{(1)} = {\tilde \bsV} + {\tilde \bsT}$ into a new
``classical part'' ${\tilde \bsV}$ given by
\be
{\tilde \bsV} \,=\, \bsV + \sum_{j=1}^{2} \lambda^j \bsT^{(1)\rm ll}_j
\end{equation}
and ${\tilde \bsT}$ contains all remaining terms. The new perturbation
satisfy the bounds ${\tilde \bsT}^{\rm ll} = {\rm O}(\lambda^3)$, ${\tilde
\bsT}^{\rm hh} = {\rm O}(\lambda)$, and  ${\tilde \bsT}^{\rm lh} = {\rm
O}(\lambda^2)$. If $\tilde \bsV$ is a classical interaction with a
sufficiently regular zero-temperature phase diagram, then Pirogov-Sinai 
techniques can be applied to study the phase diagrams of ${\tilde \bsV} + 
{\tilde \bsT}$ for sufficiently small $\lambda$ (see below).

Note that this perturbation
scheme is not only useful to analyze the low-temperature behavior of the
model.  The new ``classical part''
${\tilde \bsV}$ does not need to be classical at all. For example, see
\cite{DFFR,DFF2}, if one applies this perturbation scheme to the Hubbard model at 
half-filling, ${\tilde \bsV}$ is given by the Heisenberg model and this
gives a rigorous proof of the equivalence of both models up to 
controlled error terms.

\section{Phase diagrams, contour models, and Pirogov-Sinai theory}
\label{secPS}

A phase diagram in Thermodynamics is a partition of a space of
physical parameters in domains corresponding to phases; the free
energy varies very smoothly inside a domain.  However, first
derivatives or of higher order may have discontinuities when crossing
the boundary between two domains, and in this case one talks of {\it
phase transitions}.

The first proof of a phase transition was proposed by Peierls for the
Ising model \cite{Pei}.  It was extended by Pirogov and Sinai
\cite{PS,Sin} to situations where different phases are not related by
a symmetry. Important
extensions and simplifications of the Pirogov-Sinai theory include
Koteck\'y and Preiss \cite{KP}, Zahradn\'\i k \cite{Zah}, Bricmont
{\it et.al} \cite{BKL} and \cite{BS}, Borgs and Imbrie \cite{BI},
Borgs and Koteck\'y \cite{BK,BK2}. An exposition of the Pirogov-Sinai
theory can be found in \cite{EFS}.

Another extension of the Peierls argument was done in Fr\"ohlich and
Lieb \cite{FL} using reflection positivity \cite{FSS,DLS}.

\subsection{Phase diagrams}
\label{secphd}

We consider the Banach space $\caB_r$ of periodic interactions, with the norm defined in
\eqref{s.int}. Here $r$ is any positive number, but further assumptions (bounds for the
weights of the contours, see below) can be verified in given models only if $r$ is large
enough. To a given interaction $\bsH \in \caB_r$ and temperature $\beta$ we
associate the set of all translation invariant (or periodic) KMS
states or, equivalently \cite{Ara, Isr}, the set of all tangent
functionals to the free energy $f(\bsH)$. The set of periodic KMS
states forms a simplex, so that it is enough to describe the extremal
states, or the corresponding tangent functionals. We denote the set of
extremal states by $\caE^\beta(\bsH)$.

In order to define a phase diagram we consider a smooth
$(p-1)$-dimensional manifold on the Banach space $\caB_r$ of periodic
interactions; it is described by an application $u \mapsto
\bsH^u$, from a connected open set $\caU \subset \bbR^{p-1}$ into $\caB_r$. For
$m=1,2,3,\dots$, we introduce $E^{(m)} = \{ \bsH \in \caB_r:
|\caE^\beta(\bsH)|=m \}$; accordingly, we partition the set $\caU$ as
\be
\label{defphd}
\caU = \union_{m=1}^\infty \caU^{(m)}
\end{equation}
where $u \in \caU^{(m)}$ iff $\bsH^u \in E^{(m)}$. The decomposition \eqref{defphd} is called
the phase diagram of $\bsH^u$.

The phase diagram of $\bsH^u$, $u \in \caU \subset \bbR^{p-1}$, is
said to satisfy the {\it Gibbs phase rule} if the following conditions
hold. Here, we call ``boundary" of $\caU^{(i)}$ the set
$(\bar\caU^{(i)} \setminus \caU^{(i)}) \cap \caU$, with
$\bar\caU^{(i)}$ the closure of $\caU^{(i)}$.

\begin{itemize}
\item[(i)] $\caU = \caU^{(1)} \cup \dots \cup \caU^{(p)}$.
\item[(ii)] \begin{itemize}
\item[(a)] $\caU^{(1)}$ consists of $p$ connected components, each of which 
is a
$(p-1)$-dimensional manifold. The boundary of $\caU^{(1)}$ is
$\caU^{(2)} \cup\dots\cup
\caU^{(p)}$.
\item[(b)] $\caU^{(2)}$ consists of $\bigl(\begin{smallmatrix} p\\2
\end{smallmatrix}\bigr)$ connected components, each of which is a 
$(p-2)$-dimen\-sional
manifold. The boundary of $\caU^{(2)}$ is $\caU^{(3)} \cup\dots\cup
\caU^{(p)}$.
\item[(c)] $\caU^{(q)}$ consists of $\bigl(\begin{smallmatrix} p\\q
\end{smallmatrix}\bigr)$ connected components, each of which is a 
$(p-q)$-dimen\-sional
manifold. The boundary of $\caU^{(q)}$ is $\caU^{(q+1)} \cup\dots\cup
\caU^{(p)}$.
\item[(d)] $\caU^{(p)}$ consists of a single point $u_0$.
\end{itemize}
\end{itemize}

In other words, the phase diagram of $\bsH^u$ satisfies the Gibbs
phase rule iff it is homeomorphic to a connected, open neighborhood
$\caU'$ of the boundary of the positive octant of $\bbR^p$, in such a
way that $u_0$ is mapped onto the origin, $\caU^{(p-1)}$ is mapped
onto the union of axis $\cup_i \{ a_i>0, a_j=0, j\neq i\}$, and so
on...

Connected components of $\caU^{(1)}$ are the {\it one-phase} region,
or {\it pure phase} region, $\caU^{(2)}$ is the region of coexistence
of two phases, \dots, $\caU^{(p)}$ is the point of coexistence of all
$p$ phases.

We will call a phase diagram which satisfies the Gibbs phase rule {\it
regular} if the free energy is a real analytic function of $u$ in each
one-phase region, and if all connected components of the manifold
$\caU^{(j)}$ are smooth ($C^1$).

\subsection{Contour models}
\label{seccontmod}

A {\it contour} $\caA$ is a pair $(A,\alpha)$, where $A \subset
\bbZ^\nu$ is a finite connected set and is the {\it support} of $\caA$; to
describe $\alpha$, let us introduce the closed unit cell $C(x) \subset
\bbR^\nu$ centered at $x$, i.e.\ $C(x) = \{ y \in \bbR^\nu : |y-x|_\infty \leq \frac12 \}$.
The {\it boundary} $B(A)$ of $A \subset \bbZ^\nu$ is the union of plaquettes
\be
B(A) = \{ C(x) \cap C(y) : x \in A, y \notin A \}.
\end{equation}
The boundary $B(A)$ decomposes into connected components; each connected 
component $b$ is
given a label $\alpha_b \in \{1,\dots,p\}$, and $\alpha = (\alpha_b)$.

Let $\Lambda \subset \bbZ^\nu$ finite, with periodic boundary
conditions. A set of contours $\{\caA_1, \dots, \caA_k\}$ is {\it
admissible} iff
\begin{itemize}
\item $A_i \subset \Lambda$, and $\dist(A_i,A_j) \geq 1$ if $i\neq j$.
\item Labels $\alpha_j$ are matching in the following sense. Let $W = \Lambda 
\setminus
\cup_{j=1}^k A_j$; then each
connected component of $W$ must have same label on its boundaries.
\end{itemize}
For $j \in \{1,\dots,p\}$, let $W_j$ be the union of all connected
components of $W$ with labels $j$ on their boundaries.

For each $j \in \{1,\dots,p\}$, we give ourselves a complex function $g^{\beta,u}_j$
(``free energy of a restricted ensemble"), that is real analytic in $u
\in \caU$. We suppose that the limit $\beta\to\infty$ of
$g^{\beta,u}_j$ exists, and we write
\ba
e^u_i &= \lim_{\beta\to\infty} \Re g_i^{\beta,u}, \quad 1\leq i\leq p,\\
e_0^u &= \min_i e^u_i.
\label{deffen0}
\end{align}
We consider the partition function \eqref{deffpart} for an interaction
$\bsH^u = \bsV^u + \bsT$, where the periodic interaction $\bsT$ is a
perturbation of $\bsV^u$. We assume that the partition function can
be rewritten as
\be
Z_\Lambda^{\beta,u,\bsT} = \sum_{\{\caA_1,\dots,\caA_k\}} \prod_{j=1}^k 
w^{\beta,u,\bsT}(\caA_j)
\prod_{i=1}^p \e{-\beta g_i^{\beta,u} |W_i|},
\end{equation}
where the sum is over admissible sets of contours in
$\Lambda$.\footnote{The sum includes the case $k=0$, and the corresponding term is $\sum_{j=1}^p \e{-\beta g_i^{\beta,u}
|\Lambda|}$. It is however irrelevant, since it does not contribute to the infinite-volume
free energy \eqref{deffen}.}  The
{\it weight} $w^{\beta,u,\bsT}(\caA)$ of a contour $\caA$ is a complex
function of $\beta$, $u$, and $\bsT$, that behaves nicely for $\beta$
large and $\bsT$ in a neighborhood of 0. Precisely, we assume that
there exists a set $\caW \subset \bbR_+
\times \caB_r$, that is open and connected, and whose closure contains $(\infty,0)$;
furthermore, we suppose that for all $u \in \caU$ and
all $(\beta,\bsT) \in \caW$, and all contours $\caA$,
\begin{itemize}
\item $w^{\beta,u,\bsT}$ is periodic with period $\ell$, i.e.\ we have
$w^{\beta,u,\bsT}(\tau_a \caA) = w^{\beta,u,\bsT}(\caA)$ for all $a
\in (\ell\bbZ)^\nu$ and all $\caA$. Here $\tau_a$ is the translation
operator.
\item $|w^{\beta,u,\bsT}(\caA)| \leq \e{-\beta e_0^u |A|}
\e{-\tau|A|}$ for a large enough constant $\tau$ (depending on $\nu$, $p$, 
and $\ell$). Furthermore, $|\frac\partial{\partial u_i}
w^{\beta,u,\bsT}(\caA)| \leq \beta |A| C \e{-\beta e_0^u |A|}
\e{-\tau|A|}$ and $|\frac\partial{\partial\eta}
w^{\beta,u,\bsT+\eta\bsK}(\caA)| \leq \beta |A| C \|\bsK\|_r \e{-\beta e_0^u |A|}
\e{-\tau|A|}$ for a uniform constant $C$.
\item $\lim_{\beta\to\infty} \lim_{\bsT\to0}
w^{\beta,u,\bsT}(\caA) = 0$. This means that the weights represent the
correction to the situation $(\beta=\infty,\bsT=0)$.
\item $w^{\beta,u,\bsT}(\caA)$ is real analytic in $u$; for all 
$\bsK \in \caB_r$, $w^{\beta,u,\bsT+\eta\bsK}(\caA)$ is real analytic in
$\eta$ in a neighborhood of 0 (the neighborhood depends on $\bsK$).
\end{itemize}

Finally, the free energy is
\be
\label{deffen}
f^{\beta,u,\bsT} = -\frac1\beta \lim_{\Lambda \nearrow \bbZ^\nu} 
\frac1{|\Lambda|} \log
Z_\Lambda^{\beta,u,\bsT}.
\end{equation}
We also assume the following properties for $f^{\beta,u,\bsT}$:
\begin{itemize}
\item $f^{\beta,u,\bsT}$ is real, and concave as a function of $\bsT$;
\item whenever $\bsH^u + \bsT = \bsH^{u'} + \bsT'$, we have
\be
\label{assf}
f^{\beta,u,\bsT} = f^{\beta,u',\bsT'}.
\end{equation}
\end{itemize}
Although these properties seem difficult to verify in the
context of a contour model, they are usually clear in the original
physical model.

\subsection{The Pirogov-Sinai theory}

The results of the Pirogov-Sinai theory are usually presented in terms
of existence of many Gibbs states for a given interaction. However, it
is more convenient to think of the Pirogov-Sinai theory as to express
the free energy in a suitable form for the description of first-order
phase transitions: the free energy is given as the minimum of $C^1$
functions (``metastable free energies"), that intersect themselves by
making angles.  Hence a first-order phase transition when varying
parameters so as to cross an intersection.

The free energy at zero temperature is given by \eqref{deffen0}; in
typical situations this is the minimum over energies of some important
configurations (the ``potential ground states"). The Pirogov-Sinai
theory shows that in contour models, this structure extends at low
temperatures. In the quantum situation one is also interested in
adding a perturbation to a ``nice" model; the metastable free energies
then depend not only on $\beta$, but also on the quantum perturbation.

We claim that the Pirogov-Sinai theory allows to construct metastable free energies that
satisfy the following properties.

\vspace{2mm}
\noindent
{\bf Properties of the metastable free energies.}
\label{propmfe}
{\it
We consider a contour model that satisfies the structure described in Section
\ref{seccontmod}. Then there exist $p$ real functions $f^{\beta,u,\bsT}_i$ for
$(\beta,\bsT,u) \in \caW\times\caU$, such that
\begin{itemize}
\item[(a)] $f^{\beta,u,\bsT} = \min_i f_i^{\beta,u,\bsT}$;
\item[(b)] $\lim_{\beta\to\infty} \lim_{\bsT\to0} f_i^{\beta,u,\bsT} = e_i^u$, and
$\lim_{\beta\to\infty} \lim_{\bsT\to0} \frac\partial{\partial u_j}
f_i^{\beta,u,\bsT} =
\frac\partial{\partial u_j} e_i^u$;
\item[(c)] for all $\bsK \in \caB_r$, there exists a neighborhood $\caN_\bsK$ of 0 such that
$f^{\beta,u,\bsT+\eta\bsK}_i$ is $C^1$ as a function of $(u,\eta)$ in
$\caU\times\caN_\bsK$, and $|\frac\partial{\partial\eta} f_i^{\beta,u,\bsT+\eta\bsK}| \leq
C \|\bsK\|_r$ for a constant $C$ depending on $\nu,p,\ell$ only;
\item[(d)] $f_i^{\beta,u,\bsT}$ is a real analytic function of $u$ in $\frM_{\{i\}} = \{ u: f_i^{\beta,u,\bsT} < f_j^{\beta,u,\bsT} \,
\fall j\neq i\}$.
\end{itemize}
}
\vspace{2mm}

Notice that the point (d) implies that the free
energy $f^{\beta,u,\bsT}$ is a real analytic function of $u$ in
$\cup_i
\frM_{\{i\}}$ (which is the region of uniqueness, as will be seen below).

The proof of these properties involves the full artillery of the Pirogov-Sinai theory. The
item (c) is not really standard and may appear as superfluous technicalities, but it plays a role when
establishing the properties of the phase diagram, see Theorem \ref{thmphd} below. Since the
present paper is only aimed at studying a special class of quantum models, we content
ourselves with an outline of the proof, so as to make it plausible for readers who have
knowledge of the details of the Pirogov-Sinai theory. A review of the Pirogov-Sinai theory is expected to
appear shortly and will contain a detailed proof of these properties.

\begin{proof}[Sketch of the proof of these properties]
We heavily rely on \cite{BKU}, which itself follows
\cite{PS, Sin, Zah, BI, BK, BK2}. Our metastable free energies are
defined as the real part of the metastable free energies of
\cite{BKU}, which are complex in general.

The first step consists in defining the metastable free energies. This can be done by
introducing truncated contour activities and truncated partition functions following the
inductive procedure of \cite{BKU}, Eqs (5.6)--(5.12). One obtains metastable free energies
$f_j^{(n)}$ (that depend on $\beta,u,\bsT$). One can then prove the claims of Lemma A.1 i),
iii), iv), v), vi) of \cite{BKU}. We set then $f_j^{\beta,u,\bsT} = \lim_{n\to\infty}
f_j^{(n)}$.

At this point we have well-defined metastable free energies depending on $\beta, u$ and
$\bsT$ (that is, they are functionals on the Banach space of interactions), and the
free energy of the system is given by the minimum of the metastable free energies, as
stated in item (a). It is also clear that $\lim_{\beta\to\infty}
\lim_{\bsT\to0} f_i^{\beta,u,\bsT} = e_i^u$, and that $f_i^{\beta,u,\bsT}$ is real analytic
in $u$ on $\frM_{\{i\}}$. What remains to be done is to check differentiable properties.

For given $\bsT$ and $\bsK$, we consider $f_j^{\beta,u,\bsT+\eta\bsK}$ as a function of
$(u,\eta)$. This is a mild complication of the situation in \cite{BKU}, since the
metastable free energies here depend on $p$ parameters instead of $p-1$. One then gets the items
ii) and vii) of Lemma A.1 --- the partial derivatives with respect to $\eta$ of the truncated contour activities
and of the partition function with given external label satisfying the claims of the
lemma with a constant $C_0 \|\bsK\|_r$ instead of $C_0$.

Finally, the metastable free energies are given as convergent series of clusters of
contours, the weights of those obeying suitable bounds. This leads to item (c).
\end{proof}

We show now that these metastable free energies allow for a complete
characterization of tangent functionals, under the extra assumption
that the situation at zero temperature and without perturbation
satisfies the Gibbs phase rule in a strong sense.

The stronger condition for the Gibbs phase rule is that, for some $u_0 \in \caU$, we have
that all ``potential ground state energies'' are equal,
$e_i^{u_0} = e_j^{u_0}$ for all $i,j$, and that the matrix of derivatives
\be
\label{matrixder}
\Bigl( \frac\partial{\partial u_j} \bigl[ e_i^u - e_p^u \bigr] 
\Bigr)_{1\leq i,j \leq p-1}
\end{equation}
has an inverse that is uniformly bounded. Actually, energies $e_i^u$ may not be
differentiable; in this case, we consider the same matrix with 
$\Re g_i^{\beta,u}$ instead of $e_i^u$, and we suppose that it has an
inverse for all $\beta$ large enough, the inverse matrix being uniformly
bounded with respect to $u \in \caU$, and $\beta \geq \const$.

\begin{theorem}[Stability of the phase diagram]
\label{thmphd}

Assume that there exist metastable free energies $f^{\beta,u,\bsT}_i$,
$1\leq i\leq p$, that satisfy all points {\it (a)--(d)} of the properties above. We assume in
addition that the strong version of the
Gibbs phase rule, described above, is satisfied.

Then for $\beta$ large enough and $\|\bsT\|_r$ small enough (depending
on $p$ and on the bound of the inverse of the matrix of derivatives
\eqref{matrixder}), there exists $\caU'
\subset \caU$ such that the phase diagram for 
$\bsH^u + \bsT$, $u \in \caU'$, at inverse temperature
$\beta$, satisfies the Gibbs phase rule and is regular.
\end{theorem}

Theorem \ref{thmphd} states that there exists $u_0' \in
\caU'$ such that the set of tangent functionals to the free energy at 
$\bsH^{u_0'} + \bsT$
is a simplex with $p$ extremal points. More generally, we have the
decomposition $\caU' = {\caU'}^{(1)} \cup\dots\cup {\caU'}^{(p)}$ such that for 
$u \in {\caU'}^{(q)}$, the set of
tangent functionals at $\bsH^u + \bsT$ is a $q$-dimensional simplex.

This ``completeness" of the phase diagram was addressed in \cite{Zah}
and \cite{BW}. The approach was however different and involved
studying the Gibbs states, which is more intricate and does not easily
extend to the quantum case. It is simpler to look at tangent
functionals, and then to use existing results on their equivalence
with DLR or KMS states.

Notice that the Pirogov-Sinai theory also provides various extra
informations, such as the fact that the limit of ${\caU'}^{(q)}$, as
$\bsT\to0$ and $\beta\to\infty$, is equal to $\caU^{(q)}$. Also, the
extremal equilibrium states can be shown to be exponentially
clustering. We do not claim these properties here however, because
doing so would require extra assumptions and technicalities in the
description of the abstract contour model.

\begin{proof}[Proof of Theorem \ref{thmphd}]

Items (b) and (c) of the properties of metastable free energies (with $\eta=0$) imply that there
exists $u_0'$ such that $f^{\beta,u_0',\bsT}_i =
f^{\beta,u_0',\bsT}_j$ for all $i,j$, and that the matrix of
derivatives
\be
\label{matderfen}
\Bigl( \frac\partial{\partial u_j} \bigl[ f^{\beta,u,\bsT}_i - f^{\beta,u,\bsT}_p \bigr] 
\Bigr)_{1\leq i,j \leq p-1}
\end{equation}
has a bounded inverse, uniformly in $u$ in a neighborhood $\caU'$ of $u_0'$. Let us define
\be
\frM_i = \{ u \in \caU' : f^{\beta,u,\bsT}_i = \min_j f^{\beta,u,\bsT}_j \},
\end{equation}
and, for $Q \subset \{1,\dots,p\}$,
\be
\frM_Q = \bigcap_{i\in Q} \frM_i \setminus \bigcup_{i\notin Q} \frM_i
\end{equation}
(notice that $\frM_{\{i\}} \subsetneq \frM_i$). By the implicit
function theorem, each $\frM_Q$ is described by a $C^1$ function from
an open subset of $\bbR^{p-|Q|}$ into $\caU'$. If we set 
$\caU^{(q)}= \cup_{|Q|=q} \frM_Q$ the phase diagram satisfies the Gibbs 
phase rule, provided there are
exactly $|Q|$ tangent functionals at $\bsH^u+\bsT$ for each $u \in
\frM_Q$.

Each metastable free energy $f^{\beta,u,\bsT}_j$, 
$j\in Q$, defines a tangent functional $\alpha_j$: for all $\bsK \in \caB_r$, we set
$\alpha_j(\bsK) = \frac\partial{\partial\eta} f_j^{\beta,u,\bsT+\eta\bsK} |_{\eta=0}$.
Notice that item (c) ensures boundedness of the tangent functional.\footnote{One may wonder
whether the functional $\alpha_j$ is linear. It is actually, because $\alpha_j$ can be obtained as the limit of
linear functionals that are tangent to the free energy, uniquely defined for all points of $\frM_{\{j\}}$ --- a
region of parameters where the concave free energy has a unique tangent functional.} We show
now that these tangent functionals are linearly independent, and that any other
tangent functional is a linear combination of these ones.

We examine the manifold where $q$ phases coexist; without loss of
generality, we can choose $\tilde u \in \frM_Q$ with $Q = \{ 1, \dots,
q \}$. The determinant of
\eqref{matderfen} can be written as a linear combination of determinants of
\be
\Bigl( \frac\partial{\partial u_{k_j}} \bigl[ f^{\beta,\tilde u,\bsT}_i -
f^{\beta,\tilde u,\bsT}_q \bigr] 
\Bigr)_{1\leq i,j \leq q-1},
\end{equation}
with $k_1, \dots, k_{q-1}$ being $q-1$ different indices. Since the
determinant of
\eqref{matderfen} differs from 0, at least one of the determinants 
in the previous equation differs from 0. Without loss of generality 
we can assume that
\be
\label{submatrix}
\Bigl( \frac\partial{\partial u_j} \bigl[ f^{\beta,\tilde u,\bsT}_i 
- f^{\beta,\tilde
u,\bsT}_q \bigr]  \Bigr)_{1\leq i,j \leq q-1}
\end{equation}
is not singular.

Our analysis is local, so we can take $\tilde u=0$ and
$\bsH^u = \bsH^0 +
\sum_{j=1}^{p-1} u_j \bsK_j$. Then \eqref{assf} implies that $\alpha_j(\bsK_i) 
= \frac\partial{\partial u_i}
f^{\beta,u,\bsT}_j |_{u=0}$, and non-singularity of \eqref{submatrix} shows that
$\alpha_j$, $1\leq j\leq q$, are linearly independent. Furthermore, it also implies that
for all tangent functional $\alpha'$ the system of
equations for $\xi = (\xi_1,\dots,\xi_q)$,
\be
\alpha'(\bsK_i) = \sum_{j=1}^q \xi_j \, \alpha_j(\bsK_i), \quad i = 1,\dots,q-1,
\end{equation}
has a unique solution with $\sum_j \xi_j = 1$ . Now we consider any
$\bsK \in \caB_r$; we define $g_j(u,\eta) =
f_j^{\beta,u,\bsT+\eta\bsK}$, $1\leq j\leq q$, and
\be
g(u,\eta) = \left( \begin{matrix} g_1(u,\eta) - g_q(u,\eta) \\ \vdots \\ g_{q-1}(u,\eta) -
g_q(u,\eta) \end{matrix} \right).
\end{equation}
We have $g(0,0) = 0$, $\frac\partial{\partial u} g(0,0)$ is an
isomorphism, and $g(u,\eta)$ is a map of class $C^1$ by item
(c) of the properties of page \pageref{propmfe}.  By the implicit function theorem there exists a map
$u(\eta)$ such that $g(u(\eta),\eta)=0$. We introduce the interactions
\ba
&\bsR(\eta) = \bsK + \frac1\eta \sum_{j=1}^{q-1} u_j(\eta) \bsK_j, \\
&\bsR = \lim_{\eta\to0} \bsR(\eta) = \bsK + \sum_{j=1}^{q-1} u_j'(0) \bsK_j.
\end{align}
Then using \eqref{assf} we have
\be
f^{\beta,0,\bsT + \eta\bsR(\eta)} = f^{\beta,0,\bsT + \eta\bsR(\eta)}_1 = \dots 
= f^{\beta,0,\bsT +
\eta\bsR(\eta)}_q.
\end{equation}
Differentiating with respect to $\eta$, we obtain 
(recall that $\alpha'$ is tangent to
$f^{\beta,0,\bsT + \eta\bsR(\eta)}$ at $\eta=0$)
\be
\alpha'(\bsR) = \alpha_1(\bsR) = \dots = \alpha_q(\bsR).
\end{equation}
Then obviously $\alpha'(\bsR) = \sum_j \xi_j \alpha_j(\bsR)$, and it 
follows by linearity
of the tangent functionals that
\be
\alpha'(\bsK) = \sum_{j=1}^q \xi_j \, \alpha_j(\bsK).
\end{equation}
\end{proof}

\section{Results of the quantum Pirogov-Sinai theory}
\label{secresults}

We summarize in this section the results obtained in \cite{BKU, DFF,
DFFR, KU}, and in the present paper. All results concern the situation
where the interaction has the form $\bsH = \bsV+\bsT$, where $\bsV$ is a 
classical interaction satisfying the standard Pirogov-Sinai framework, and 
$\bsT$ is a small perturbation. The temperature will be assumed to be
small. The results however split into four classes, according to
whether we use the perturbation methods of \cite{DFFR} (Section 2.2), 
and whether we include high temperature expansions to analyze phases at 
intermediate temperatures.

In this section, we implicitely assume all properties of the metastable free energies, see
page \pageref{propmfe}, to be valid --- without these properties the statements
below would not include completeness, i.e.\ we could not ascertain to have identified {\it
all} the periodic Gibbs states
of the systems.

\subsection{Quantum perturbation of classical model with finitely many 
ground states}

In this case the classical interaction $\bsV$ has finitely 
many ground states and the phase diagram of $\bsV +\bsT$ is, 
at low temperatures and for sufficiently small $\bsT$ a small 
deformation of the zero temperature phase diagram of $\bsV$. 
The extension of the Pirogov-Sinai theory to this class of quantum systems 
goes back to \cite{Pir} and was proved in \cite{BKU,DFF}.

\subsubsection{(a) Structure:} We denote by 
$\Omega = \{ 1, \dots, M \}^{\bbZ^\nu}$ the space of classical
configurations; the dimension $\nu$ of the physical space is always
supposed to be bigger or equal to 2.  The interaction has the form 
$\bsH=\bsV+\bsT$, where $\bsV$ is a block interaction and is diagonal 
with respect to the basis of classical configurations: if 
$A = U(x)\equiv\{y\,:\, |y-x|_\infty \leq R\}$ for some 
$x \in \bbZ^\nu$,
\be
V_A {\ket e_\omega} = \Phi_x(\omega_{U(x)}) {\ket e_\omega},
\end{equation}
and $V_A=0$ if there is no $x$ with $U(x) = A$. The function $\Phi_x$ 
depends on $\bsmu \in \caU \subset \bbR^{p-1}$,
and we assume that its derivatives $\frac\partial{\partial\mu_j} \Phi_x(\omega_{U(x)})$ are
bounded uniformly in $x,\bsmu,\omega,j$.

A finite set $G = \{ g^{(1)}, \dots, g^{(p)} \} \subset \Omega$ 
of periodic configurations is given, that
contains all ground states of $V$ for all $\bsmu$ (see below the
precise assumption). We write $G_A = \{ g_A : g \in G \}$. We suppose
that $\Phi_x(g_{U(x)})$ is independent of $x$, for all $g \in G$, and
this value is denoted by $e^\bsmu_g$ (this is the mean energy of the
configuration $g$).

\subsubsection{(b) Assumptions:}
\begin{itemize}
\item[{\bf (A1)}] A gap separates the excitations: for all $\omega_{U(x)}
\notin G_{U(x)}$,
$$
\Phi_x(\omega_{U(x)}) - \min_{g \in G} \Phi_x(g_{U(x)}) \geq \Delta
$$
(uniformly in $\bsmu$).
\item[{\bf (A2)}] The zero temperature phase diagram is (linearly) regular: 
there is $\bsmu_0 \in \caU$ such that $e^{\bsmu_0}_g = e^{\bsmu_0}_{g'}$ for 
all $g,g' \in G$, and the inverse of the matrix of derivatives $M^\bsmu_G$, 
see \eqref{matrixder}, is uniformly bounded.
\end{itemize}

\subsubsection{(c) Properties of Gibbs states:}

\begin{theorem}
\label{thmbasic}
Assume {\bf (A1)} and {\bf (A2)} hold true.  There exist $\beta_0, c <
\infty$ (depending on $\nu, R, p, M$ and on the
periods of $\{ g^{(j)} \}$ and $\bsH$ only) such that if $\beta \Delta
\geq \beta_0$ and $\| \bsT \|_c / \Delta \leq 1$, the phase
diagram of the quantum model satisfies the Gibbs phase rule and is regular 
in a neighborhood $\caU' \subset \caU$ of $\bsmu_0$.

In the single phase region, \ie if $\bsmu \in \frM^\beta(\{ g \})$, 
the KMS state $w^{\beta,\bsmu,\bsT}(\cdot)$ is close to the ground state $g$: 
for all $K \in \caO_A$, 
$\lim_{\beta \rightarrow \infty, \|\bsT\|_r \rightarrow 0} 
w^{\beta,\bsmu,\bsT}(K) \,=\, \langle e_g |K|e_g \rangle$. 
\end{theorem}

The condition $\|\bsT\|_c/\Delta \leq 1$ means that $\bsT$ is a
perturbation with respect to $\bsV$; $c$ plays the role of the
perturbative parameter: from the definition \eqref{s.int} of the norm
$\|\cdot\|_c$, $\|T_A\|$ must be very small if $c$ is very large.

The proof of this theorem follows from \cite{BKU, DFF}.

\subsection{Models with infinite degeneracy}

Consider a model whose classical part has infinitely many
ground states, and a perturbation which lifts this degeneracy
completely. The pertubation methods of \cite{DFFR}(see section 2.2)
permits in certain cases to analyze this by constructing an equivalent
interaction with a new classical part which has finitely many
ground states. In this case the new perturbation has a slightly more
complicated form than in Section 4.1 and the following theorem deals
with this situation.  This situation was considered in
\cite{DFFR} (for a different approach see \cite{KU}).

\subsubsection{(a) Structure:}

The space of classical configurations is again $\Omega = \{ 1, \dots,
M \}^{\bbZ^\nu}$. We consider two sets $G,D \subset \Omega$, with $D
\subset G$ finite, $D = \{ d^{(1)}, \dots, d^{(p)} \}$ is a finite set 
of periodic configurations; $G$ may be infinite and will represent the 
configurations of low energy.  For $A \subset \bbZ^\nu$, the Hilbert space 
$\caH_A$ has the following
decomposition $\caH_A = \caH_A^\low \oplus \caH_A^\high$ where
$\caH_A^\low$ is the subspace spanned by the low energy configurations
$g_A \in G_A$. The interaction has the form $\bsH = \bsV+\bsT$, where
$\bsV$ is a classical block interaction with uniformly bounded
derivatives $\frac\partial{\partial\bsmu_j} \Phi_x(\omega_{U(x)})$,
and $\bsT$ is a perturbation that is submitted to some restrictions,
see the assumptions below.

\subsubsection{(b) Assumptions:}
\begin{itemize}
\item[{\bf (B1)}] A gap separates high and low energies: for all $\omega_{U(x)} 
\notin G_{U(x)}$,
$$
\Phi_x(\omega_{U(x)}) - \max_{g \in G} \Phi_x(g_{U(x)}) \geq \Delta_0.
$$
\item[{\bf (B2)}] Gap with the ground states: we assume that
$\Phi_x(d_{U(x)})$ is
independent of $x$ for $d \in D$, and for all $\omega_{U(x)} \notin D_{U(x)}$,
$$
\Phi_x(\omega_{U(x)}) - \min_{d \in D} \Phi_x(d_{U(x)}) \geq \Delta
$$
(and we assume that $\Delta \leq \Delta_0$).
\item[{\bf (B3)}] The perturbation may be decomposed $\bsT = \bsK + \bsK' +
\bsK''$; for all $A$,
\ba
&K_A' \caH_A^\low = 0, &K_A' \caH_A^\high \subset \caH_A^\high; \nn\\
&K_A'' \caH_A^\low \subset \caH_A^\high, &K_A'' \caH_A^\high \subset
\caH_A^\low \nn
\end{align}
(there is no assumption on $\bsK$).\footnote{Motivation comes from
\eqref{structLS}.  It is however slightly more general, and it is just what
is required in the proof of Theorem
\ref{thmLS}.}
\item[{\bf (B4)}] The zero temperature phase diagram is (linearly) regular,
\ie all
energies $e^\bsmu_d$ are equal for some $\bsmu_0 \in \caU$, and the matrix
$M^\bsmu_D$ [see
\eqref{matrixder}] has a uniformly bounded inverse.
\end{itemize}

\subsubsection{(c) Properties of Gibbs states:}

\begin{theorem}
\label{thmLS}
Assume {\bf (B1)}--{\bf (B4)} hold true.  There exist $\beta_0, c <
\infty$ (depending on $\nu, R, p, M$ and on the periods of $\{
d^{(j)} \}$ and $H$ only) such that if $\beta \Delta \geq \beta_0$,
$\| \bsK \|_c / \Delta \leq 1$, $\| \bsK' \|_c / \Delta_0 \leq 1$, $\| \bsK''
\|_c / \Delta_0 \leq 1$ the phase diagram of the quantum model satisfies the 
Gibbs phase rule and is regular in $\caU' \subset \caU$, 
$\caU' \ni \bsmu_0$.  

In the single phase region, \ie if $\bsmu \in \frM^\beta(\{ d \})$, 
the KMS state $w^{\beta,\bsmu,\bsT}(\cdot)$ is close to the ground state $d$: 
for all $K \in \caO_A$, 
$\lim_{\beta \rightarrow \infty, \|\bsT\|_r \rightarrow 0} 
w^{\beta,\bsmu,\bsT}(K) \,=\, \langle e_d |K|e_d \rangle$. 
\end{theorem}

The proof of this theorem is given in \cite{DFFR}.  A
somewhat different method yielding similar results has been developed 
later in
\cite{KU}.

\subsection{Combined high and low temperature expansions}
\label{secintemp}

Here we consider models whose classical part $\bsV$ has partially 
ordered ground states, typically described by periodic configurations of holes 
and particles but still with infinite degeneracy due to, e.g., degeneracy 
of the spin at each site. Together with the quantum perturbation the 
system may have a continuous symmetry. 
We will suppose that the temperature is low and, in addition, 
that $\beta \| \bsT \|_c$ is actually small (\ie the temperature
is large compared to $\bsT$) and we will prove that in this case one phase 
corresponds to each  periodic configuration of holes and particles and that 
in this phase the spin degrees of freedom are in a disordered phase. 
This situation has many similarities with
that of \cite{BKL}, and could be called ``a theory of restricted
ensembles in quantum lattice systems''.

\subsubsection{(a) Structure:} As before, let $\Omega = \{ 1, \dots, M
\}^{\bbZ^\nu}$. Intermediate temperature phases will be characterized 
by ``motives'' giving partial information on the underlying configurations.  
In order to describe this, we consider a partition of $\{ 1, \dots, M \}$:
\be
\label{partition}
\{ 1, \dots, M \} = \bigcup_{j=1}^N I_j \quad \text{with } I_i \cap I_j =
\emptyset.
\end{equation}
We denote $\caN_\Lambda = \{ 1, \dots, N \}^\Lambda$ (and $\caN \equiv
\caN_{\bbZ^\nu}$). For $n \in \caN$, we write $\Omega_n = \{ \omega
\in \Omega : \omega_x \in I_{n_x} \;\forall x \}$. Let $G = \{
g^{(1)}, \dots, g^{(p)} \} \subset \caN$ be a finite set of periodic 
configurations; this is the set of motives and a pure phase will be 
associated with each of these configurations. We write 
$\Omega_G = \cup_{g \in G} \Omega_g$.

The interaction has the form $\bsH=\bsV+\bsT$, where $\bsV$ is a classical 
block
interaction with uniformly bounded derivatives w.r.t.\ $\bsmu$, and $\bsT$ is 
a perturbation.  We introduce restricted
partition functions for each $g \in G$: let
\be
Z^g_\Lambda = \sum_{\omega_\Lambda \in \Omega_{g,\Lambda}} \e{-\beta
\sum_{x, U(x) \subset
\Lambda} \Phi_x(\omega_{U(x)})}
\end{equation}
and
\be
h_g^{\beta,\bsmu} = -\frac1\beta \lim_{\Lambda \nearrow \bbZ^\nu}
\frac1{|\Lambda|} \log Z^g_\Lambda.
\end{equation}
The ground energies are $e^\bsmu_g = \lim_{\beta \to \infty}
h_g^{\beta,\bsmu}$, $g \in G$.

\subsubsection{(b) Assumptions:}
\begin{itemize}
\item[{\bf (C1)}] For all configurations $\omega_{U(x)} \notin
\Omega_{G,U(x)}$, we have
$$
\Phi_x(\omega_{U(x)}) - \min_{\omega' \in \Omega_G} \Phi_x(\omega_{U(x)}')
\geq \Delta.
$$
Moreover, we assume that
$$
\min_{\omega_{U(x)} \in \Omega_{g,U(x)}} \Phi_x(\omega_{U(x)}) = e^\bsmu(g)
$$
independently of $x$, for all $g \in G$.
\item[{\bf (C2)}] We need a condition that ensures that no phase transition 
takes place in a restricted ensemble $\Omega_g$; in other words, spatial 
correlations should decay quickly enough. The following condition is 
stronger, and amounts to saying that there is {\it no} correlation 
between different sites. For all $g \in G$, we suppose that there exists 
an on-site interaction $\Phi^g$ such that for all $x$:
$$
\Phi_x(\omega_{U(x)}) = \Phi_x^g(\omega_x)
$$
for all $\omega \in \Omega_g$.
\item[{\bf (C3)}] The zero temperature phase diagram is regular with
$e^{\bsmu_0}_g =
e^{\bsmu_0}_{g'}$, $g,g' \in G$, for some $\bsmu_0 \in \caU$, and the
matrix $M_G^\bsmu$, see \eqref{matrixder}, has a uniformly bounded inverse.
\footnote{If $\{e_g^\bsmu \}$ are not $C^1$, we consider the matrix of 
derivatives of $h_g^{\beta,\bsmu}$ for $\beta$ large; it must have an 
inverse that is bounded uniformly w.r.t.\ $\bsmu$ and
large $\beta$.}
\end{itemize}

\subsubsection{(c) Gibbs states at intermediate temperature:}

\begin{theorem}
\label{thmintemp}

Assume {\bf (C1)}--{\bf (C3)} hold true. There exist $\beta_0, c <
\infty$ (depending on $\nu, R, p, M$ and on the periods of $\{ g^{(j)}
\}$ and $\bsH$ only) such that if $\beta_0 \leq \beta \Delta < \infty$
and $\beta \| \bsT \|_c \leq 1$, the phase diagram satisfies the Gibbs phase 
rule and is regular in $\caU' \subset \caU$,  $\caU' \ni \bsmu_0$.  

In the single phase region, \ie if $\bsmu \in \frM^\beta(\{ g \})$, 
the KMS state $w^{\beta,\bsmu,\bsT}(\cdot)$ is close to the motive $g$: 
for all $K \in \caO_A$,  
$\lim_{\beta \rightarrow \infty, \|\bsT\|_r \rightarrow 0} 
w^{\beta,\bsmu,\bsT}(K) \,=\, 
({\rm Tr}(P_A))^{-1} {\rm Tr}( K P_A)$, where $P_A$ is the projection 
given by 
$\sum_{\omega_A \in \Omega_{g,A}} \ket{ e_{\omega_A}}  \bra{ e_{\omega_A}}$.
\end{theorem}

\subsubsection{Remark:} It follows from our assumptions that $\bsT$ is
small compared to
$\bsV$; more precisely, $\| \bsT \|_c / \Delta \leq 1/\beta_0$.

This theorem is actually a consequence of Theorem \ref{thmLSintemp} below,
see the remark on page \pageref{remproofthm}.

\subsection{Infinite degeneracy, high and low temperature expansions}

Here we consider systems where phases result from subtle
interplay between potential and kinetic energy, combining the effect 
described in Section 4.2 and 4.3.  The quantum perturbation lifts partially 
the degeneracy of the classical interaction, leading at intermediate 
temperatures, to spatially ordered phases.   
Hereafter we describe the general framework in a rather abstract way; it
will be illustrated in Section \ref{secexample}, and the reader may gain
better understanding by working out a concrete application.

\subsubsection{(a) Structure:}
The space of classical configurations is $\Omega = \{ 1, \dots, M
\}^{\bbZ^\nu}$; we consider a partition like in \eqref{partition} and
define similarly $\caN$ and $\Omega_n$. We consider a (possibly
infinite) set $G \subset \caN$ that represents low energy
configurations; the Hilbert spaces decompose in the following way:
$\caH_A = \caH_A^\low \oplus \caH_A^\high$, where $\caH_A^\low$ is the
subspace spanned by the low-energy configurations $g_A \in G_A$.  The
interaction has the form $\bsH=\bsV+\bsT$; $\bsV$ is a block
interaction with uniformly bounded derivatives
$\frac\partial{\partial\bsmu_j} \Phi_x(\omega_{U(x)})$; the
perturbation $\bsT$ decomposes further $\bsT = \bsK+\bsK'+\bsK''$; we
shall require different assumptions on $\bsK,\bsK',\bsK''$, motivated
by the perturbation theory of Section 2.2.

We suppose that a finite set $D = \{ d^{(1)}, \dots, d^{(p)} \}
\subset G$ is given, that corresponds to possible ground states. For
each $d \in D$, we define the corresponding restricted partition
function
\be
Z_\Lambda^d = \sum_{\omega_\Lambda \in \Omega_{d,\Lambda}} \e{-\beta
\sum_{x, U(x) \subset \Lambda} \Phi_x(\omega_{U(x)})}
\end{equation}
and the corresponding restricted free energy
\be
h_d^{\beta,\bsmu} = -\frac1\beta \lim_{\Lambda \nearrow \bbZ^\nu}
\frac1{|\Lambda|} \log Z_\Lambda^d,
\end{equation}
and $e^\bsmu_d = \lim_{\beta \to \infty} h_d^{\beta,\bsmu}$.

\subsubsection{(b) Assumptions:}

\begin{itemize}
\item[{\bf (D1)}] A gap separates high and low energies: for all
$\omega_{U(x)} \notin \Omega_{G,U(x)}$,
$$
\Phi_x(\omega_{U(x)}) - \max_{\omega' \in \Omega_G} \Phi_x(\omega_{U(x)}')
\geq \Delta_0.
$$
\item[{\bf (D2)}] Gap with the ground states: for all $\omega_{U(x)} \notin
\Omega_{D,U(x)}$,
$$
\Phi_x(\omega_{U(x)}) - \min_{\omega' \in \Omega_D} \Phi(\omega_{U(x)}')
\geq \Delta.
$$
\item[{\bf (D3)}] For all $d \in D$, there exists an on-site interaction 
$\Phi^d$ such that for all $\omega \in \Omega_d$ and all $x$,
$$
\Phi_x(\omega_{U(x)}) = \Phi_x^d(\omega_x).
$$
Moreover, we suppose that
$$
\min_{\omega_x \in I_{d_x}} \Phi_x^d(\omega_x) = e^\bsmu_d
$$
independently of $x$.
\item[{\bf (D4)}] The quantum perturbation $\bsT = \bsK+\bsK'+\bsK''$ has
same properties as in {\bf (B3)}, with respect to the decomposition into 
low and high energy states.
\item[{\bf (D5)}] There is $\bsmu_0 \in \caU$ such that $e^{\bsmu_0}_d =
e^{\bsmu_0}_{d'}$, $d,d' \in D$, and the matrix of derivatives
\eqref{matrixder} has a uniformly bounded inverse (see the footnote of {\bf (C3)} if
$e_d^\bsmu$ is not differentiable).
\end{itemize}

\subsubsection{(c) Properties of Gibbs states:}

\begin{theorem}
\label{thmLSintemp}

Assume {\bf (D1)}--{\bf (D5)} hold true. There exist $\beta_0, c <
\infty$ (depending on $\nu, R, p, M$ and on the periods of $\{ d^{(j)}
\}$ and $\bsH$ only) such that if $\beta_0 \leq \beta \Delta < \infty$,
$\beta \| \bsK \|_c \leq 1$, $\| \bsK' \|_c / \Delta_0 \leq 1$, $\| \bsK''
\|_c / \Delta_0 \leq 1$, and $\beta \| \bsK'' \|_c^2 / \Delta_0 \leq 1$,
the phase diagram satisfies the Gibbs phase rule and is regular in 
an open set $\caU' \subset \caU$ that contains $\bsmu_0$.  

In the single phase region, \ie if $\bsmu \in \frM^\beta(\{ d \})$, 
the KMS state $w^{\beta,\bsmu,\bsT}(\cdot)$ is close to the motive $d$: 
for all $K \in \caO_A$,  
$\lim_{\beta \rightarrow \infty, \|\bsT\|_r \rightarrow 0} 
w^{\beta,\bsmu,\bsT}(K) \,=\, 
({\rm Tr}(P_A))^{-1} {\rm Tr}( K P_A)$, where $P_A$ is the projection 
given by 
$\sum_{\omega_A \in \Omega_{d,A}} \ket{ e_{\omega_A}}  \bra{ e_{\omega_A}} $.
\end{theorem}

This theorem follows from the contour representation obtained in Section
\ref{secexp}, together with the Pirogov-Sinai theory.

\subsubsection{Remarks:}
\label{remproofthm}
1. Theorem \ref{thmintemp} is an immediate consequence of Theorem
\ref{thmLSintemp}. Indeed, we clearly recover the setting of Section
\ref{secintemp} by choosing $G = \Omega$ (\ie all configurations have
low energy), and $\bsK' = \bsK'' = 0$.

2. These two theorems also generalize results of \cite{Uel}: they can be
applied to the Hubbard model
\be
H = -t \sumtwo{\neighbours xy}{\sigma = \uparrow,\downarrow} (c^\dagger_{x\sigma}
c_{y\sigma} + {\rm h.c.}) + U \sum_x n_{x\uparrow} n_{x\downarrow},
\end{equation}
to show that the high temperature phase extends to
$$
\bigl\{ (\beta,t,U): \beta t \text{ small} \bigr\} \quad \text{ and } \quad
\bigl\{ (\beta,t,U): \beta t^2/U \text{ small} \bigr\}
$$
(standard high temperature expansions apply when both $\beta t$ and $\beta U$ are small).

\section{Example: Extended Hubbard model}
\label{secexample}

This is a Hubbard model where particles interact among each other when
their distance is smaller or equal to 1.  Explicitly,
\be
H_\Lambda = -t \sumtwo{\neighbours xy \subset \Lambda}{\sigma = \uparrow,\downarrow} 
(c^\dagger_{x\sigma} c_{y\sigma} + {\rm h.c.}) + U \sum_{x \in \Lambda} n_{x\uparrow} n_{x\downarrow} 
+ W \sum_{\neighbours xy \subset \Lambda} n_x n_y - \mu \sum_{x \in \Lambda} n_x.
\end{equation}
Here, $c^\dagger_{x,\sigma}, c_{x\sigma}$ are creation, annihilation,
operators of a fermion of spin $\sigma$ at site $x$; $\neighbours xy$
stands for a set of nearest neighbor sites; $n_{x\sigma} =
c^\dagger_{x\sigma} c_{x\sigma}$ is the number of particles of spin
$\sigma$ at $x$ (it has eigenvalues 0 and 1); $n_x = n_{x\uparrow} +
n_{x\downarrow}$ is the total number of particles at $x$. The
coefficient $t$ represents the hopping, and will be taken to be small
compared to the nearest-neighbor repulsion $W$; $\mu$ is the chemical
potential. The classical limit $t\to0$ was studied in \cite{Jed,BJK}. The stability of the
chessboard phase $\frM_{(0,2)}$ (see below) with small $t$ is a straightforward application
of \cite{DFF}; a later study devoted to it is \cite{BK3}.

We start by analyzing the classical interactions.  The configuration space is 
$\Omega =\{0,\uparrow,\downarrow,2\}^{\bbZ^\nu}$ and the corresponding classical
interaction can be written as (taking $R = \tfrac12$)
\be
\Phi_x(\omega_{U(x)}) = \frac U{2^\nu} \sum_{y \in U(x)}
\delta_{\omega_y,2} + \frac W{2^{\nu-1}} \sum_{\neighbours yz \subset U(x)}
q_y q_z - \frac\mu{2^\nu} \sum_{y \in U(x)} q_y.
\end{equation}
Here we introduced $q_y \in \{0,1,2\}$:
\be
q_y = \begin{cases} 0 & \text{if } \omega_y = 0 \\ 1 & \text{if } \omega_y
= \uparrow \text{ or } \omega_y = \downarrow \\ 2 & \text{if } \omega_y =
2.  \end{cases}
\end{equation}

The interaction can also be written as a sum over pairs of n.n.\ sites;
this simplifies the analysis of the zero temperature phase diagram, and the
search for symmetries (see below).  This pair interaction is given by
\be
\Phi_{\neighbours xy}(q_x,q_y) = \frac U{2\nu} (\delta_{q_x,2} +
\delta_{q_y,2}) + W q_x q_y - \frac\mu{2\nu} (q_x + q_y).
\end{equation}

This model has a hole-particle symmetry.  Introducing the unitary operator
$U$ such that $U c^\dagger_{x\sigma} U^{-1} = c_{x\sigma}$ and $U
c_{x\sigma} U^{-1} = c^\dagger_{x\sigma}$, we see that $U T_\Lambda U^{-1}
= T_\Lambda$.  As for the potential, the effect of the symmetry can be
exhibited by considering classical configurations; defining $q_x' = 2 -
q_x$, and $\mu' = U + 4\nu W - \mu$, we easily check that
\be
\Phi_{\neighbours xy}^{\mu'}(q_x',q_y') = \Phi_{\neighbours
xy}^\mu(q_x,q_y) + C
\end{equation}
where $C = -U/\nu - 4W + 2\mu/\nu$ does not depend on $(q_x,q_y)$.  As a
result, the phase diagrams $(U,\mu)$ are symmetric along the line
\be
\label{hpsym}
\mu = \frac U2 + 2\nu W,
\end{equation}
for any temperature.

The zero temperature phase diagrams with $t=0$ are depicted in
\fig\ref{phdextHub0}, in both cases $W<0$ and $W>0$.

\bfig
\centerline{$\begin{matrix} \text{\epsfxsize=60mm \epsffile{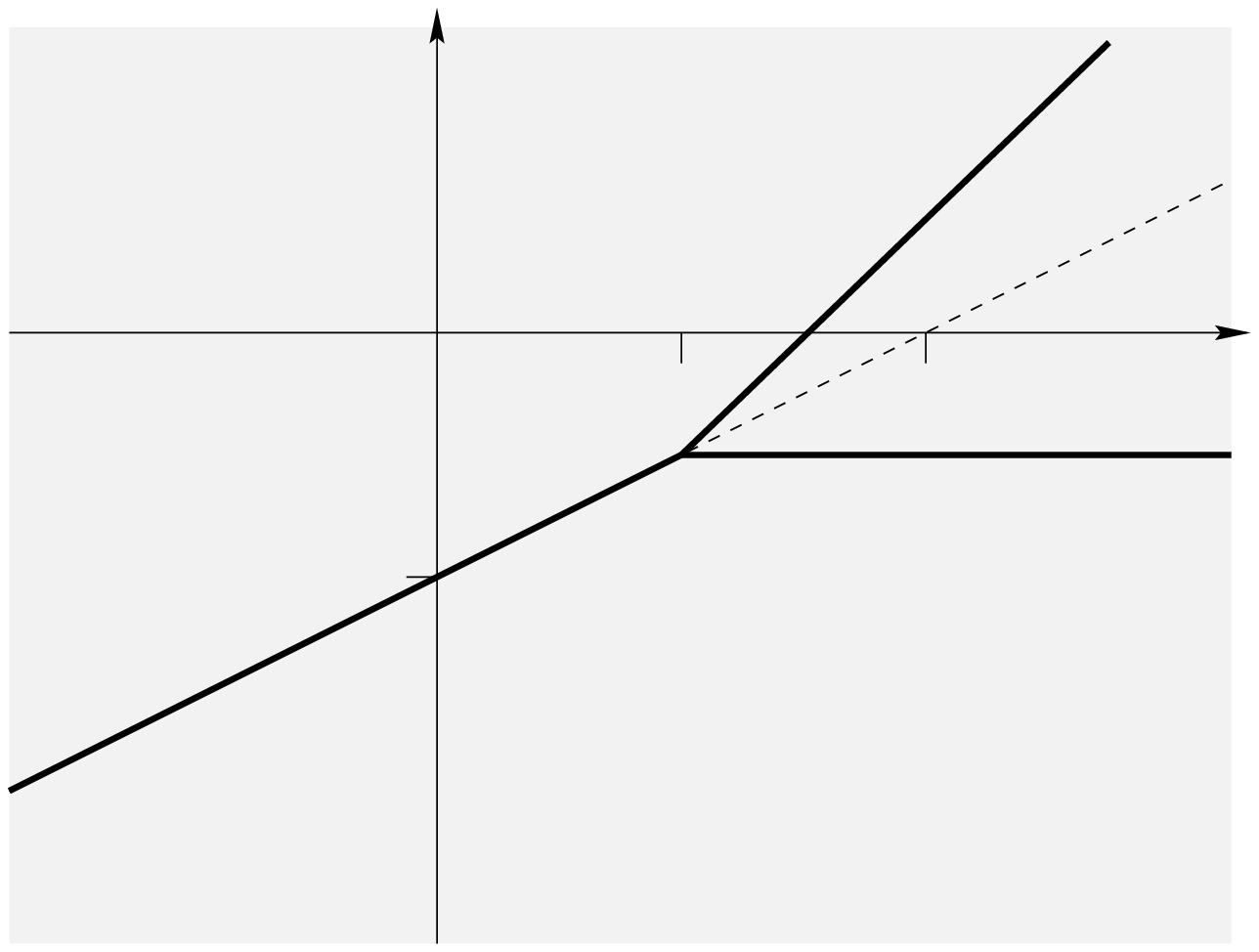}}
\\ (a) \end{matrix} \hspace{15mm} \begin{matrix} \text{\epsfxsize=60mm
\epsffile{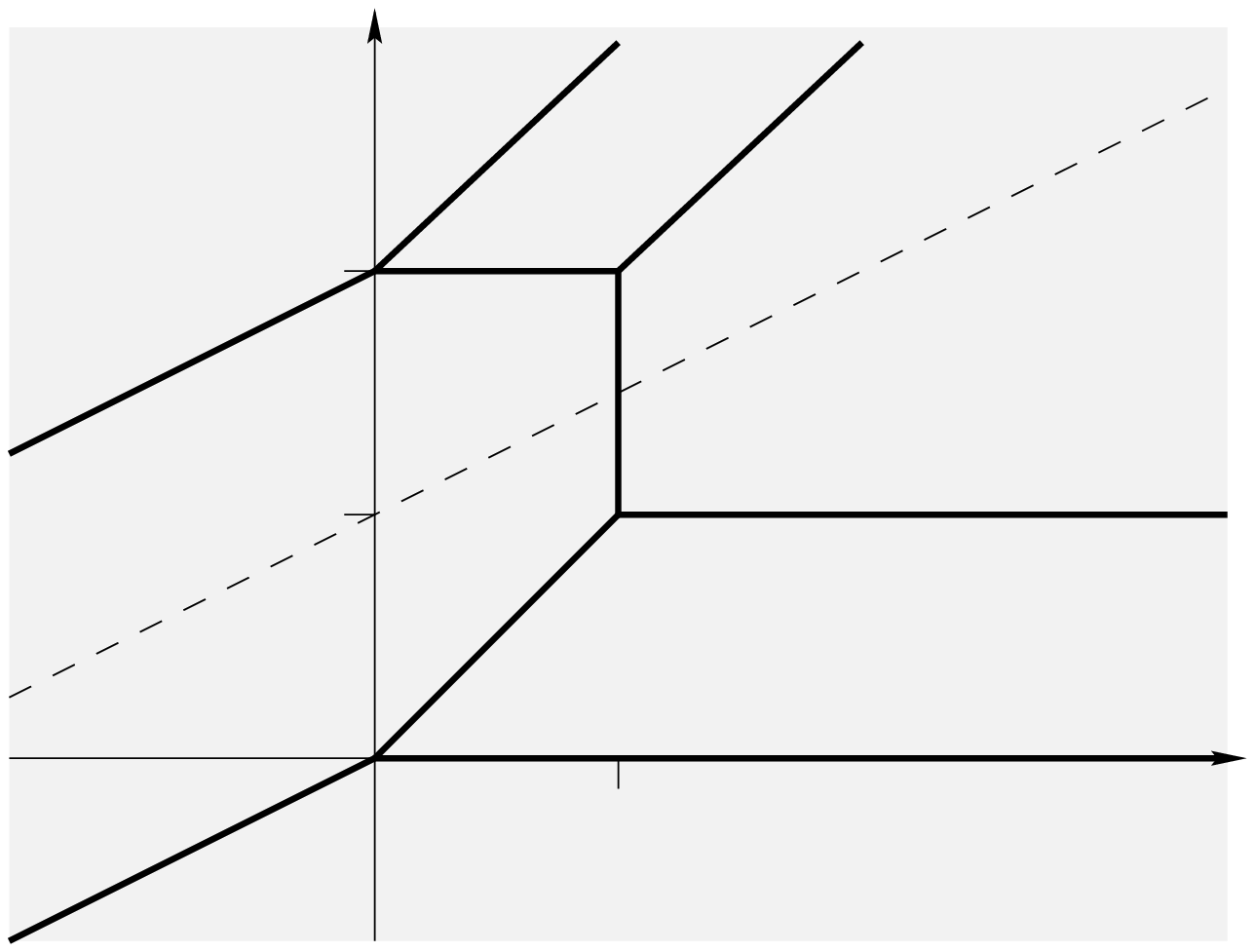}} \\ (b) \end{matrix}$}
\figtext{
\writefig	-0.7	3.5	{\footnotesize $\tfrac U{\nu|W|}$}
\writefig	-4.5	5.1	{\footnotesize $\tfrac\mu{\nu|W|}$}
\writefig	-3.57	3.2	{\footnotesize 2}
\writefig	-2.43	3.2	{\footnotesize 4}
\writefig	-5.2	2.4	{\footnotesize -2}
\writefig	-3.8	4.3	{\footnotesize $\frM_2$}
\writefig	-1.5	4.5	{\footnotesize $\frM_1$}
\writefig	-3.3	1.7	{\footnotesize $\frM_0$}
\writefig	6.8	1.4	{\footnotesize $\tfrac U{\nu W}$}
\writefig	2.7	5.1	{\footnotesize $\tfrac\mu{\nu W}$}
\writefig	2.2	2.65	{\footnotesize 2}
\writefig	2.2	3.85	{\footnotesize 4}
\writefig	3.65	1.15	{\footnotesize 2}
\writefig	1.4	4.3	{\footnotesize $\frM_2$}
\writefig	5.3	3.6	{\footnotesize $\frM_1$}
\writefig	4.5	1.0	{\footnotesize $\frM_0$}
\writefig	1.1	2.7	{\footnotesize $\frM_{(0,2)}$}
\writefig	4.5	2.0	{\footnotesize $\frM_{(0,1)}$}
\writefig	3.35	4.4	{\footnotesize $\frM_{(1,2)}$}
}
\caption{Zero temperature phase diagrams of the extended Hubbard model,
$(a)$ when $W<0$ and $(b)$ when $W>0$.  The dashed line represents the
hole-particle symmetry, see \eqref{hpsym}.}
\label{phdextHub0}
\end{figure}

In the case $W<0$, it decomposes into three domains $\frM_0$, $\frM_1$, and
$\frM_2$; $\frM_0$ and $\frM_2$ have a unique translation invariant ground
state with respectively 0 and 2 particles at each site.  In $\frM_1$, any
configurations with one particle per site is a ground state; there is
degeneracy $2^{|\Lambda|}$ since each particle has spin $\uparrow$ or
$\downarrow$.

The situation $W>0$ presents a richer structure with six domains.  Domains
$\frM_0$, $\frM_1$ and $\frM_2$ have same features as with attractive n.n.\
interactions.  In between appear now domains $\frM_{(0,2)}$,
$\frM_{(1,2)}$ and $\frM_{(0,1)}$.  $\frM_{(0,2)}$ consists in two ground
states, the two chessboard configurations with alternatively 0 and 2
electrons per site.  $\frM_{(0,1)}$ has $2 \cdot 2^{\frac12 |\Lambda|}$
ground states of the chessboard type, one sublattice being empty, while the
other has exactly one particle of spin $\uparrow$ or $\downarrow$;
$\frM_{(1,2)}$ is similar, with 2 particles per site on one sublattice and
one on the other.

We are interested in the case where the temperature is small, but bigger
than 0, and with small hopping.  The phase diagrams for large $\beta$ and
small $\beta t$ are presented in \fig\ref{phdextHub}.

\bfig
\centerline{$\begin{matrix} \text{\epsfxsize=60mm \epsffile{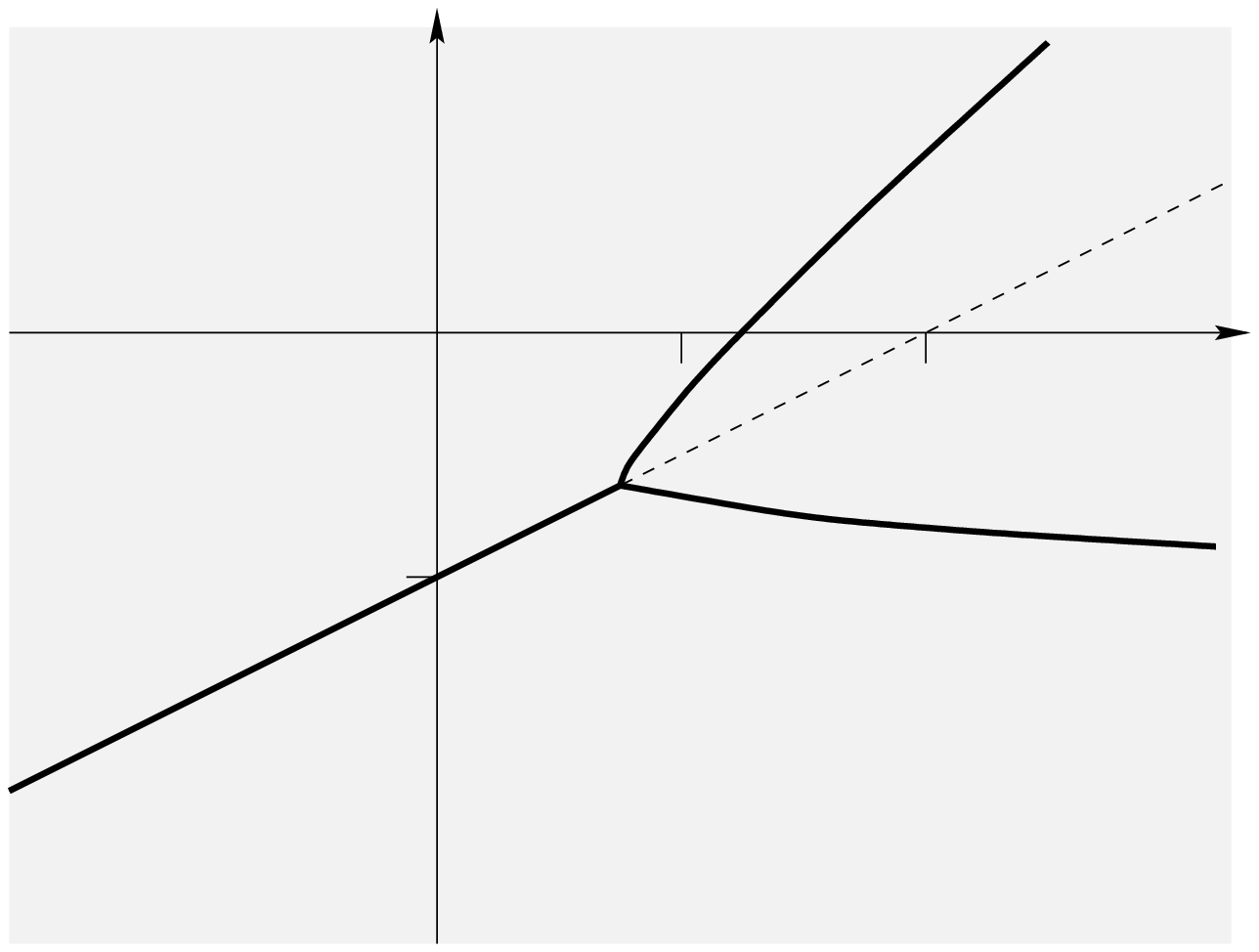}}
\\ (a) \end{matrix} \hspace{15mm} \begin{matrix} \text{\epsfxsize=60mm
\epsffile{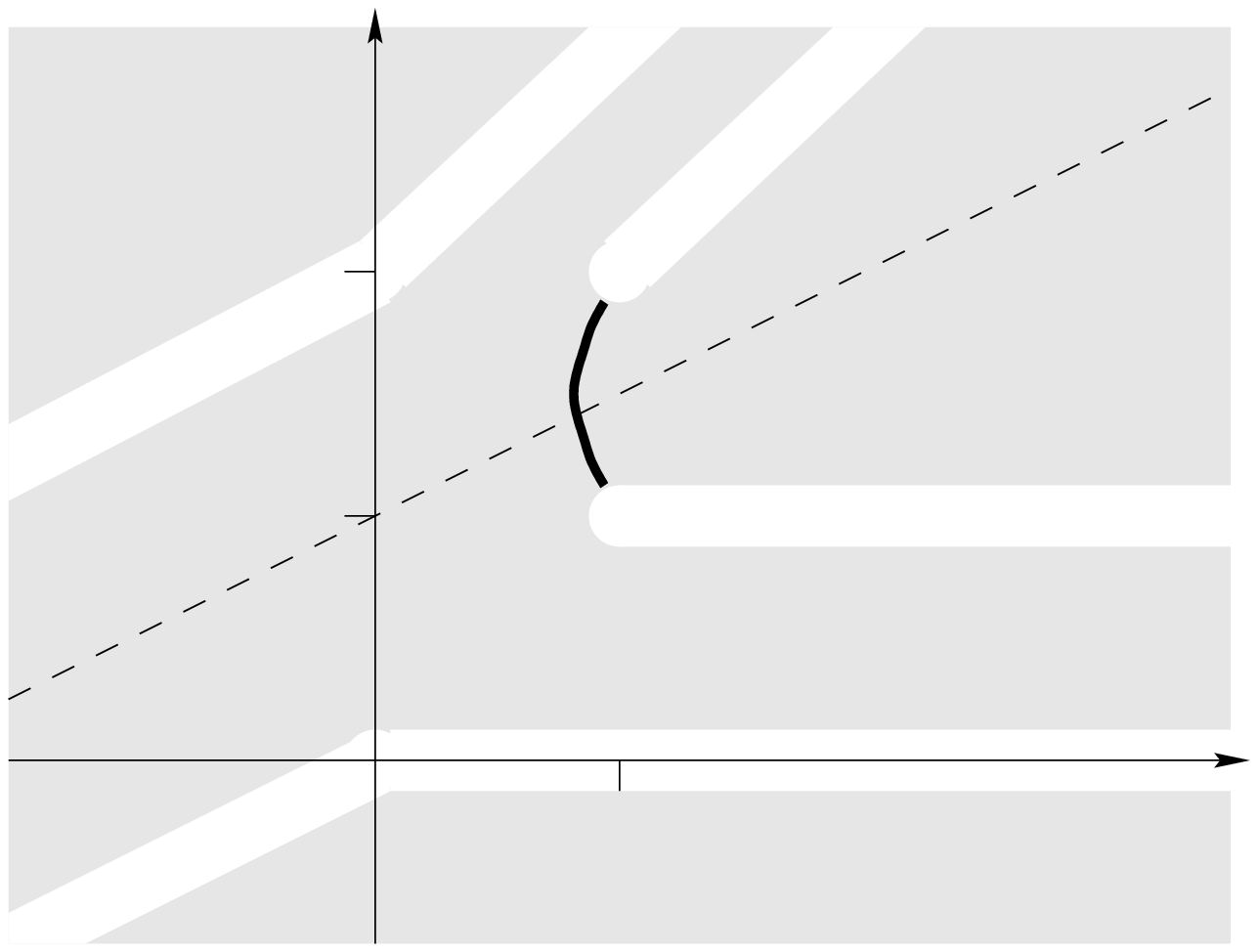}} \\ (b) \end{matrix}$}
\figtext{
\writefig	-0.7	3.5	{\footnotesize $\tfrac U{\nu|W|}$}
\writefig	-4.5	5.1	{\footnotesize $\tfrac\mu{\nu|W|}$}
\writefig	-3.57	3.7	{\footnotesize 2}
\writefig	-2.43	3.2	{\footnotesize 4}
\writefig	-5.2	2.4	{\footnotesize -2}
\writefig	-3.9	4.3	{\footnotesize $\frM_2^{\beta,t}$}
\writefig	-1.5	4.35	{\footnotesize $\frM_1^{\beta,t}$}
\writefig	-3.3	1.6	{\footnotesize $\frM_0^{\beta,t}$}
\writefig	6.8	1.4	{\footnotesize $\tfrac U{\nu W}$}
\writefig	2.7	5.1	{\footnotesize $\tfrac\mu{\nu W}$}
\writefig	2.2	2.65	{\footnotesize 2}
\writefig	2.2	3.85	{\footnotesize 4}
\writefig	3.65	1.15	{\footnotesize 2}
\writefig	1.3	4.3	{\footnotesize $\frM_2^{\beta,t}$}
\writefig	5.3	3.6	{\footnotesize $\frM_1^{\beta,t}$}
\writefig	4.5	0.9	{\footnotesize $\frM_0^{\beta,t}$}
\writefig	2.8	2.2	{\footnotesize $\frM_{\text{cb}}^{\beta,t}$}
}
\caption{Phase diagrams of the extended Hubbard model at intermediate
temperature and with small hopping, (a) when $W<0$ and (b) when $W>0$. 
Bold lines denote first-order phase transitions.  White is the region
$\caP_\epsilon$ that resists rigorous investigations, where second-order
transitions are expected.}
\label{phdextHub}
\end{figure}

In the case $W<0$, all three domains survive at low temperature and
with $t \neq 0$; a first-order phase transition occurs when crossing
the border between any two domains. The point $(\frac U{\nu W} = 2,
\frac\mu{\nu W} = 1)$ belongs to $\frM_1^{\beta,t}$: this phase has
residual entropy (it also has more quantum fluctuations,
although this has much less effect). The Gibbs
state corresponding to the domain $\frM_1^{\beta,t}$ is
thermodynamically stable and exponentially clustering. The restriction
to intermediate temperatures ($\beta t \leq \varepsilon$) is
important, because, for $\nu \geq 3$, a phase transition is expected
when the temperature decreases, leading to an antiferromagnetic phase,
that breaks both symmetries of translations and of rotations of the
spins.

The phase diagram at finite $\beta$ and non zero $t$ is especially
interesting for $W>0$.  There are not six, but only four domains
$\frM_0^{\beta,t}$, $\frM_1^{\beta,t}$, $\frM_2^{\beta,t}$ and
$\frM_{\text{cb}}^{\beta,t}$; see \fig\ref{phdextHub}.  Indeed, the three
domains corresponding to chessboard phases have merged into a single domain
(this was first understood and proven in \cite{BJK} in absence of hopping). 
The free energy is real analytic in the whole domain
$\frM_{\text{cb}}^{\beta,t}$.  The transition between $\frM_2^{\beta,t}$
and $\frM_{\text{cb}}^{\beta,t}$ is presumably second-order, but our
results do not cover the intermediate region between these domains.  The
boundary between $\frM_{\text{cb}}^{\beta,t}$ and $\frM_1^{\beta,t}$
contains a part where a first-order phase transition occurs, that can be
rigorously described.  Crossing the boundary elsewhere presumably results
in a second-order transition.  Due to the thermal fluctuations, the segment
from (2,2) to (2,4) belongs to $\frM_1^{\beta,t}$.

Our results for this model are summarized in the next two theorems.

\begin{theorem}[Hubbard model with attractive n.n.\ interactions]
\label{thmeHn}
Let $\nu\geq2$. There exist constants $\beta_0 < \infty$ and
$\varepsilon_0 > 0$ (depending on $\nu$) such that the phase diagram
$(U,\mu)$ for $\beta |W| \geq \beta_0$ and $\beta t
\leq \varepsilon_0$ is regular; domains $\frM_a^{\beta,t}$, $a \in
\{0,1,2\}$ satisfy
$\lim_{\beta\to\infty} \lim_{t\to0} \frM_a^{\beta,t} = \frM_a$. If
$(U,\mu)$ belongs to a unique $\frM_a^{\beta,t}$, there is a unique
Gibbs state. Furthermore, the density of the system is close to $a$,
$$
\bigl| \expval{n_x} - a \bigr| \leq \varepsilon(\beta,t),
$$ for all $x$.  $\varepsilon(\beta,t)$ can be made arbitrarily small by
taking $\beta$ large and $t$ small.
\end{theorem}

In order to describe the situation $W>0$ we first introduce the region of
the phase diagram $\caP_\epsilon$ where we have no results.  Let
\be
\caL = \Bigl[ \bigl( \frM_{(0,2)} \cup \frM_{(1,2)} \cup \frM_{(0,1)}
\bigr) \bigcap \bigl( \frM_0 \cup \frM_1 \cup \frM_2 \bigr) \Bigr]
\setminus \bigl[ \frM_{(0,2)} \cap \frM_1 \bigr],
\end{equation}
and for $\epsilon>0$,
\be
\caP_\epsilon = \bigcup_{(U,\mu) \in \caL} \caB_\epsilon(U,\mu)
\end{equation}
where $\caB_\epsilon(U,\mu)$ is the open ball of radius $\epsilon$
centered on $(U,\mu)$. We restrict our considerations to the
complement of $\caP_\epsilon$.

\begin{theorem}[Hubbard model with n.n.\ repulsions]
\label{thmeHp}

Let $\nu\geq2$ and $\epsilon>0$. There exist constants $\beta_0 <
\infty$ and $\varepsilon_0 > 0$ (depending on $\nu$ and $\epsilon$)
such that if $\beta_0 \leq \beta W < \infty$ and $\beta t \leq
\varepsilon_0$, we have the decomposition
$$
\caP_\epsilon^\compl = \frM_0^{\beta,t} \cup \frM_1^{\beta,t} \cup \frM_2^{\beta,t} \cup
\frM_{\rm cb}^{\beta,t},
$$
and
\begin{itemize}
\item[(i)] $\frM_0^{\beta,t} \subset \frM_0$, $\frM_2^{\beta,t} \subset \frM_2$,
$\frM_1^{\beta,t}$ ($\not\subset \frM_1$) are domains with a unique
Gibbs state. Densities are close to 0, 2, 1 respectively in the sense
\be
\begin{cases} \expval{n_x} \leq \varepsilon(\beta,t) & \text{in }
\frM_0^{\beta,t} \\ \expval{n_x} \geq 2 - \varepsilon(\beta,t) & \text{in }
\frM_2^{\beta,t} \\ |\expval{n_x} - 1| \leq \varepsilon(\beta,t) & \text{in
} \frM_1^{\beta,t} \end{cases}
\end{equation}
with $\varepsilon(\beta,t)$ arbitrarily close to 0 if $\beta$ is large and
$t$ small.
\item[(ii)] $\frM_{\rm cb}^{\beta,t} \subset \frM_{(0,2)} \cup
\frM_{(1,2)} \cup
\frM_{(0,1)}$ is a domain with two extremal Gibbs states of the chessboard
type. The free energy is a real analytic function of $\beta$
and $\mu$ in the domain $$
\bigl\{ (\beta,\mu) : \beta_0/W \leq \beta \leq \varepsilon_0/t \text{ and
} (U,\mu) \in \frM_{\rm cb}^{\beta,t} \bigr\}.  $$
\item[(iii)] $\frM_{\rm cb}^{\beta,t} \cap \frM_1^{\beta,t}$ is a line
of first-order phase transition, with exactly two three extremal states.
\end{itemize}
\end{theorem}

\subsubsection{Remarks:}
The proofs of Theorems \ref{thmeHn} and \ref{thmeHp} use Theorem
\ref{thmintemp}. But using Theorem \ref{thmbasic}, one could establish
stability of domains $\frM_0, \frM_2, \frM_{(0,2)}$ for all $\beta|W|
\geq \beta_0$, without the restriction that the temperature be not too
small. Another possible improvement, for $U,W > 0$, would use Theorem
\ref{thmLSintemp} to replace the condition $\beta t \leq
\varepsilon_0$ by $\beta t^2 /U \leq \varepsilon_0$. The later
clearly allows lower temperatures.\footnote{Furthermore, the
restriction to intermediate temperatures arises because of possible
antiferromagnetism due to ``quantum fluctuations'' of strength
$t^2/U$; it should be stable for $\beta t^2 /U > \text{const}$;
therefore this new condition is qualitatively correct.}

\section{Combined high-low temperature expansions}
\label{secexp}

In this section we simultaneously perform a low and a high temperature
expansion.  The temperature is low, in such a way that excitations above
the low energy states ($\caH_\Lambda^\low$) are rare.  At the same time,
the temperature is high relatively to the quantum perturbations $\bsK$ and
$\bsK''$.  These expansions allow to write the partition functions
as one of a {\it contour model}, that can be treated by the
Pirogov-Sinai theory, see Section \ref{seccontmod}.

We rewrite the quantum model as a contour model, by making a mixed low and
high temperature expansion (Section
\ref{secDuh}); we define suitable weights, so that the partition function
takes the form required in Section
\ref{seccontmod}.  Section \ref{secbounds} is devoted to proving that the
weights are small
compared to their size.  Finally, we explain in Section \ref{secrest} how
other requirements of Section \ref{seccontmod} are fulfilled.

\subsection{Expansion of the partition function}
\label{secDuh}

Our intention is to expand in $\bsK + \bsK' + \bsK''$; in order to
simplify the notation, we introduce $\bsB = (B,i)$, $B \subset
\bbZ^\nu$, $i = 1,2,3$, and we write $K_B = T_\bsB$ with $\bsB =
(B,1)$, $K_B' = T_\bsB$ with $\bsB = (B,2)$, and $K_B'' = T_\bsB$ with
$\bsB = (B,3)$. We refer to $\bsB$ as a {\it transition}.

Using Duhamel's formula, we obtain
\bm
\label{Duhamexp}
\Tr \e{-\beta H_\Lambda} = \Tr \e{-\beta \sum_{B \subset \Lambda} V_B}
+ \sum_{m \geq 1} \sum_{\bsB_1, \dots,
\bsB_m} \sum_{\omega^1_\Lambda, \dots, \omega^m_\Lambda}
\int_{0<\tau_1<...<\tau_m<\beta} \dd\tau_1 \dots \dd\tau_m \\
\e{-\tau_1 \sum_{x \in \Lambda} \Phi_x(\omega^1_{U(x)})} \bra{\omega^1_\Lambda}
T_{\bsB_1} \ket{\omega^2_\Lambda} \e{-(\tau_2-\tau_1) \sum_{x \in
\Lambda} \Phi_x(\omega^2_{U(x)})} \dots \\
\dots \bra{\omega_\Lambda^m} T_{\bsB_m}
\ket{\omega_\Lambda^1} \e{-(\beta-\tau_m) \sum_{x \in
\Lambda} \Phi_x(\omega^1_{U(x)})} .
\end{multline}

At this point, it is natural to define the supports of contours as all
sites that belong to $\cup_j B_j$, or for which there exists
$\omega^j$ such that $\omega_{U(x)}^j \notin
\Omega_{D,U(x)}$. But two technical difficulties arise: $d^{(1)}, \dots, d^{(p)}$ are
periodic rather than translation invariant; and the weight of a
contour should not depend on the configuration outside of its support
(but it may depend on the labeling $\alpha$).  The later difficulty
is specific to systems with phases given by a restricted ensemble
instead of a single configuration. To account for these difficulties,
we introduce a partition of the lattice into cubes of size $\ell$,
where $\ell$ is the lcm of the periods of $\{ d^{(i)} \}$ (considering
all spatial directions).

Let $\bar B = \cup_{x\in B} U(x)$; we define excited cubes.
\begin{itemize}
\item A cube $C$ is {\it quantum excited} if there is $\bsB_i$ such that $C
\cap \bar B_i
\neq \emptyset$.
\item Otherwise, it is {\it classically excited} if there is $\omega^j$ and
$x \in C$ such that $\omega^j_{U(x)} \notin \Omega_{D,U(x)}$.
\end{itemize}
Consider the set $\caQ$ of quantum excited cubes, the set $\caE$ of
classically excited cubes, and the set $\caN$ of cubes that are neighbors
of $\caQ\cup\caE$ (two cubes $C \neq C'$ are neighbors iff there exist
$x\in C$ and $y\in C'$ with $|x-y|_\infty=1$).  Connected components of
$\caQ\cup\caE\cup\caN$ form the supports of the contours.  Connected
components of the complement of $\caQ\cup\caE\cup\caN$ are characterized by
a configuration $d \in D$, and this information may be stored in the
labeling $\alpha$.  The union of all components corresponding to $d$ is
denoted $W_d$.  Then
\be
\Lambda = \caQ\cup\caE\cup\caN \cup (\cup_{d \in D} W_d),
\end{equation}
see \fig\ref{figcontours} for illustration.  $W_d$ is a union of cubes,
each cube $C$ contributing in \eqref{Duhamexp} by a factor [we use {\bf
(D3)}]
\be
\sum_{\omega_C \in \Omega_{d,C}} \e{-\beta \sum_{x \in C}
\Phi_x^d(\omega_x)} = \e{-\beta
h_d^{\beta,\bsmu} \ell^\nu}.
\end{equation}

\bfig
\epsfxsize=80mm
\centerline{\epsffile{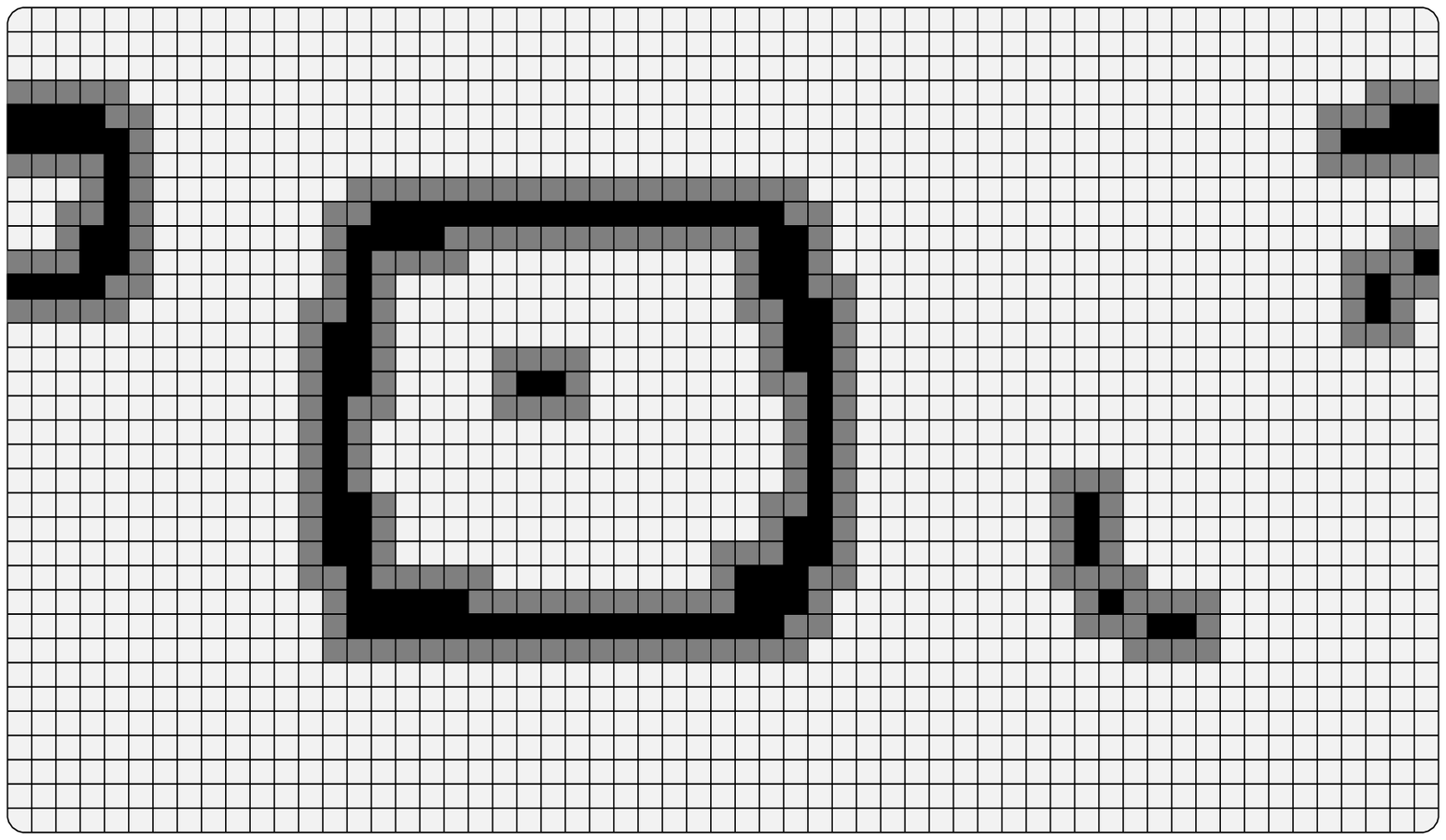}}
\figtext{
\writefig   4.1  0.5  {$\Lambda$}
\writefig   -0.5 2.9  {$\caA_1$}
\writefig   0.8  3.9  {$\caA_2$}
\writefig   3.0  3.8  {$\caA_3$}
\writefig   2.9  1.5  {$\caA_4$}
}
\caption{The space $\Lambda$ is divided into cubes; contours are formed by
excited cubes
(in black) and by their neighbors.  There are four contours on this
picture.}
\label{figcontours}
\end{figure}

Summing first over admissible sets of contours $\{\caA_1,\dots,\caA_k\}$,
we can rewrite \eqref{Duhamexp} in the following way,
\bm
\Tr \e{-\beta H_\Lambda} = \sum_{\{\caA_1,\dots,\caA_k\}} \Bigl[ \prod_{d
\in D}
\sum_{\omega_{W_d} \in \Omega_{d,W_d}} \e{-\beta \sum_{x \in W_d}
\Phi_x^d(\omega_x)} \Bigr] \\
\prod_{j=1}^k \sum_{\caQ\subset A} \sum_{m \geq 0} \sumtwo{\bsB_1, \dots,
\bsB_m}{\bar B_i
\subset \caQ} \sum_{\omega_{A_j}^1, \dots,
\omega_{A_j}^m} \int_{0<\tau_1<...<\tau_m<\beta} \dd\tau_1 \dots \dd\tau_m
\\
\e{-\tau_1 \sum_{x \in A_j} \Phi_x(\omega_{U(x)}^1)} \bra{\omega_{A_j}^1}
T_{\bsB_1}
\ket{\omega_{A_j}^2} \dots \bra{\omega_{A_j}^m} T_{\bsB_m}
\ket{\omega_{A_j}^1}
\e{-(\beta-\tau_m) \sum_{x \in A_j} \Phi_x(\omega_{U(x)}^1)}.
\end{multline}
We used here the fact that the contribution of different contours
factorizes.  There are several restrictions to the sums over transitions
$\{\bsB_i\}$ and configurations $\{\omega_{A_j}^i\}$: each cube of $\caQ$
is intersected by at least one $\bar B_i$; $\{\omega_{A_j}^i\}$ are
compatible with the labeling $\alpha_j$; excited cubes of $A_j \setminus
\caQ$ do not touch the boundary of $\caA_j$; and non excited cubes in $A_j$
have at least one neighbor that is excited.  In the last line appears
configuration $\omega^j_{U(x)}$ with $U(x) \cap W_d \neq \emptyset$, hence
depending on $\omega_{W_d}$.  However, in such a case $x$ belongs to a cube
that is not excited, so that $\omega^j_{U(x)}
\in \Omega_{d,U(x)}$. From {\bf (D3)} we can substitute $\Phi_x(\omega^j_{U(x)})$ with
$\Phi_x^d(\omega_x^j)$, which does not depend any more on the configuration
outside the support of the contour.\footnote{This is why cubes that are
neighbors of excited cubes need to be considered as part of contours.} Then
we obtain
\be
\Tr \e{-\beta H_\Lambda} = \sum_{\{\caA_1,\dots,\caA_k\}} \prod_{d \in D}
\e{-\beta
h_d^{\beta,\bsmu} |W_d|} \prod_{j=1}^k z(\caA_j)
\end{equation}
where the sum is over admissible sets of contours, and $z(\caA)$ is the
weight of the contour $\caA$.  The explicit expression for $z(\caA)$ looks
rather tedious, but the main point is to establish the properties of
Section \ref{seccontmod}.  The expression of $z(\caA)$ is
\bm
\label{defweight}
z(\caA) = \sum_{m \geq 0} \sum_{\caQ \subset A} \sumtwo{\bsB_1, \dots,
\bsB_m}{\bar B_i \subset \caQ} \sum_{\omega^1_A, \dots, \omega^m_A}
\int_{0<\tau_1<...<\tau_m<\beta} \dd\tau_1 \dots \dd\tau_m \\
\e{-\tau_1 \sum_{x \in A} \Phi_x(\omega^1_{U(x)})} \bra{\omega^1_A}
T_{\bsB_1} \ket{\omega^2_A} \dots \bra{\omega_A^m} T_{\bsB_m}
\ket{\omega_A^1} \e{-(\beta-\tau_m) \sum_{x \in A} \Phi_x(\omega^1_{U(x)})},
\end{multline}
with some restrictions on the sums over $\{\bsB_i\}$ and $\{\omega_A^i\}$,
see above.

\subsubsection{Remark:} we constructed contours out of cubes, while the
supports of
contours in Section \ref{seccontmod} are any connected sets.  There is no
contradiction, if we define the weight $z(\caA)$ to be 0 if the support of
$\caA$ is not a union of cubes.

\subsection{Bounds for the weights of the contours}
\label{secbounds}

We turn to the proof of the exponential decay of the weight of contours, as
required in Section \ref{seccontmod}.  We give the following ``space-time''
interpretation to the collection of sums and integrals in
\eqref{defweight}: we view $(\bar B_j,\tau_j)$ as a subset of $A
\times [0,\beta]_\per$, with periodic boundary conditions along the
``vertical'' interval $[0,\beta]$.  Furthermore, to each ``time'' $\tau \in
[0,\beta]_\per$ corresponds the configuration $\omega^j$ for which
$(\tau_{j-1},\tau_j] \ni \tau$.  We define
\ba
&\bsB = \bigcup_{j=1}^m \bar B_j \times \{\tau_j\} \bigcup \caQ \times \{0\};
\nn\\
&\bsE = \bigcup_{j=1}^{m+1} E(\omega^j_\caQ) \times [\tau_{j-1}, \tau_j], \quad |\bsE| =
\sum_{j=1}^{m+1} |E(\omega^j_\caQ)| (\tau_j - \tau_{j-1}), \nn
\end{align}
(with $\tau_0 \equiv 0$, $\tau_{m+1} \equiv \beta$, and $\omega^{m+1}
\equiv \omega^1$).  Here, we set $\bar B = \cup_{x \in B} U(x)$, and
$E(\omega_\caQ) = \{ x\in\caQ: \omega_{U(x)} \notin
\Omega_{G,U(x)} \}$.

>From assumptions {\bf (D1)} and {\bf (D2)} we can bound
\bm
|z(\caA)| \leq \e{-\beta e_0^\bsmu |A|} \sum_{m \geq 0} \sum_{\caQ\subset
A} \sumtwo{\bsB_1, \dots,
\bsB_m}{\bar B_i \subset \caQ} \sum_{\omega^1_A, \dots, \omega^m_A}
\int_{0<\tau_1<...<\tau_m<\beta} \dd\tau_1 \dots \dd\tau_m \\
\e{-\beta\Delta |\caE|/\ell^\nu}
\e{-\Delta_0 |\bsE|} \prod_{j=1}^m
\| T_{\bsB_j} \|,
\end{multline}
where the sums over $\{\bsB_j\}$ and $\{\omega_A^j\}$ satisfy the
restrictions explained above.

We view each $\bar B_j$ as a connected subset of $\bbR^{\nu+1}$ (one can
\eg add links between nearest neighbors).  Then $\bsB\cup\bsE$ is a subset
of $\bbR^{\nu+1}$ made out of vertical segments and horizontal sets.  We
consider connected components of $\bsB \cup
\bsE$.  For a connected component with $m$ horizontal sets and $m' \geq
m-1$ vertical segments, we deleted $m'-m+1$ of the latters, in such a way
that the component remains connected.  One of these components contains
$\caQ \times \{0\}$, possibly with extra vertical segments and horizontal
sets.  Other components have $m$ horizontal sets and $m-1$ vertical
segments.  Because of the structure ${\bf (D4)}$, or ${\bf (B3)}$,
components not linked with $\caQ \times \{0\}$, either consists in a single
transition of type $\bsK$, or include at least two transitions of type
$\bsK$ or $\bsK''$.

\bfig
\epsfxsize=80mm
\centerline{\epsffile{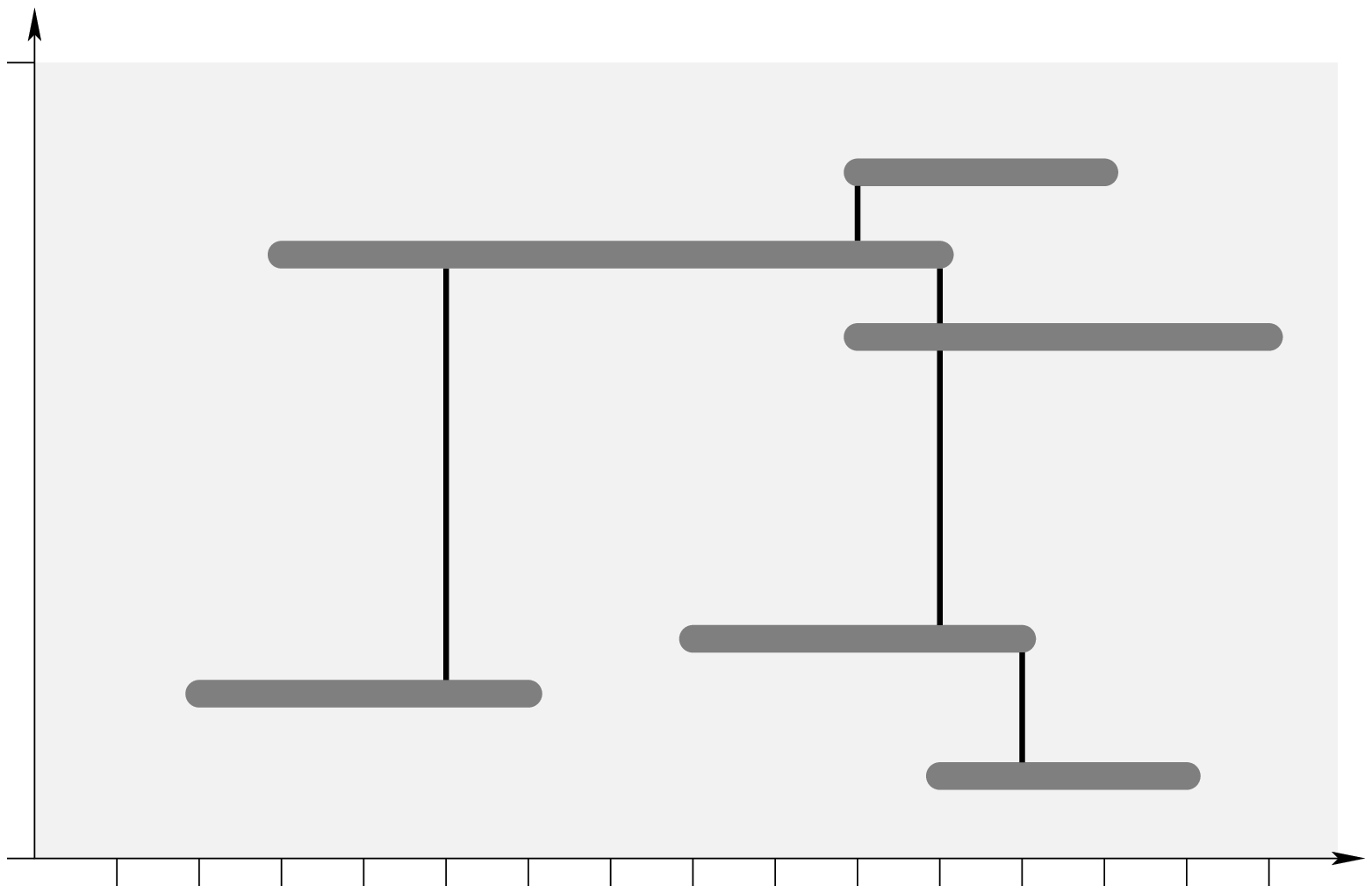}}
\figtext{
\writefig   4.1  0.4  {$A$}
\writefig   -4.4  5.25  {$\beta$}
\writefig   -4.3  0.55  {0}
\writefig   0.45  2.15  {$\bar B_j$}
}
\caption{A ``gather'' with 6 transitions and 5 vertical segments.}
\label{figgather}
\end{figure}

A connected object with $m$ horizontal sets, and $(m-1)$ vertical segments
that end on the horizontal sets, is called a {\it gather} and is denoted by
the letter $\caG$.  It is illustrated in \fig \ref{figgather}.  We
introduce the following sets of gathers:
\begin{itemize}
\item $\frG_m$: gathers with $m$ horizontal sets, one containing the origin
$\{x=0\} \times
\{\tau=0\}$; $\frG = \frG_\infty$.
\item $\frG'$: gathers of $\frG_1$ that consist in a unique transition of
type $\bsK$.
\item $\frG_m''$: gathers of $\frG_m$, with at least two transitions of
type $\bsK$ or
$\bsK''$; $\frG = \frG_\infty''$.
\end{itemize}

The connected component of $\bsB\cup\bsE$ that contains $\caQ \times \{0\}$
can be viewed as a set of gathers, each gather being connected to $\caQ
\times \{0\}$ by a vertical segment.

Since a choice of $\{\bsB_i\}$ and $\{\omega_\caQ^i\}$ leads to a set of
gathers, we obtain a bound by first integrating over sets of gathers, then
summing over compatible space-time configurations $\bsomega_A$, and
choosing which gathers are linked to $\caQ \times \{0\}$.  Therefore
\be
|z(\caA)| \leq \e{-\beta e_0^\bsmu |A|} \sum_{k\geq0} \frac1{k!}
\int \dd\caG_1 \dots \dd\caG_k \sum_{\bsomega_A} \sum_{\text{links}}
\e{-\beta\Delta |\caE|/\ell^\nu}
\e{-\Delta_0 |\bsE|} \prod_{j=1}^k \Bigl( \prod_{\bsB\in\caG_j} \|T_\bsB\|
\Bigr).
\end{equation}
The shortcut $\int\dd\caG$ means a sum over the number $m$ of transitions,
a sum over transitions $\bsB_1, \dots, \bsB_m$, an integral over ordered
times $\tau_1, \dots,
\tau_m$, and a sum over $(m-1)$ vertical segments that link $\bigl\{ \bar
B_i \times \{\tau_i\}
\bigr\}$ together.

We define
\be
\tilde z(\caG) = \e{-\Delta_0 |\caG|} \prod_{\bsB\in\caG} \|T_\bsB\|
\e{2\nu \ell^\nu (\log M +
\tau) |\bar B|},
\end{equation}
where $|\caG|$ is the total length of the vertical segments of $\caG$.
If $\beta\Delta/\ell^\nu \geq 2\nu (\log M + \tau)$, we can write
\bm
z(\caA)| \e{\tau|A|} \leq \e{-\beta e_0^\bsmu |A|} \sum_{k\geq0}
\frac{(2|\caQ|)^k}{k!}
\Bigl( \int_0^\beta \dd\tau \e{-\Delta_0\tau} \int_\frG \dd\caG \tilde
z(\caG) \Bigr)^k \\
\sum_{k
\geq 0} \frac{|\caQ|^k}{k!} \Bigl( \int_0^\beta \dd\tau
\int_{\frG'\cup\frG''} \dd\caG \tilde z(\caG)
\Bigr)^k,
\label{boundzA}
\end{multline}
where the first sum corresponds to the number of gathers linked to $\caQ
\times \{0\}$, and the second sum is the number of independent gathers. 
The shortcut $\int_\frG \dd\caG$ is identical to $\int\dd\caG$, except for
the absence of an integral over $\tau_1$, which is set to 0; integrals over
$\frG'$ and $\frG''$ are similar.

One easily obtains an upper bound for the gathers with a unique transition:
\be
\int_{\frG'} \dd\caG \tilde z(\caG) = \sum_{B, \bar B \ni 0} \|K_B\|
\e{2\nu \ell^\nu (\log M + \tau)
|\bar B|} \leq (2R+1)^\nu \|\bsK\|_c
\end{equation}
with $c = 2\nu \ell^\nu (2R+1)^\nu (\log M + \tau)$; this is smaller than
$\Delta_0$ if $c$ is large enough in the assumptions of Theorem
\ref{thmLSintemp}.

For general gathers, we proceed by induction. First,
\be
\int_{\frG_1} \dd\caG \tilde z(\caG) \leq (2R+1)^\nu \bigl( \|\bsK\|_c +
\|\bsK'\|_c + \|\bsK''\|_c
\bigr).
\end{equation}
Next, we use the recursion inequality:
\bm
\int_{\frG_m} \dd\caG \tilde z(\caG) \leq \sum_{\bsB, \bar B \ni 0}
\|T_\bsB\| \e{2\nu
\ell^\nu (\log M + \tau) |\bar B|} \\
\sum_{k\geq0} \frac1{k!} \Bigl( 2\sum_{y \in \bar B}
\int_0^\beta \dd\tau \e{-\Delta_0 \tau} \int_{\frG_{m-1,0}} \dd\caG \tilde
z(\caG)
\Bigr)^k.
\end{multline}
Integrating over $\tau$, and since $\|\bsK^{\boldsymbol\cdot}\|_c/\Delta_0
\leq 1$ for a large enough $c$, we get
\ba
\int_{\frG_m} \dd\caG \tilde z(\caG) &\leq \sum_{\bsB, \bar B \ni 0}
\|T_\bsB\| \e{2\nu
\ell^\nu (\log M + \tau) |\bar B|} \e{2|\bar B|} \nn\\
&\leq (2R+1)^\nu \bigl( \|\bsK\|_c + \|\bsK'\|_c + \|\bsK''\|_c
\bigr).
\end{align}
This holds independently of $m$.  This allows to estimate the integral over
gathers that contain at least two transitions of type $\bsK$ or $\bsK''$. 
Let $\frG_m' \subset \frG_m$ be gathers with at least one transition of
type $\bsK$ or $\bsK'$.  One easily obtains
\be
\int_{\frG_m} \dd\caG \tilde z(\caG) \leq (2R+1)^\nu \bigl( \|\bsK\|_c +
\|\bsK''\|_c
\bigr).
\end{equation}
Then the integral over gathers with two transitions of type $\bsK$ or
$\bsK''$ can be done by integrating first on the time for such a
transition, then over vertical segments and gathers at their ends, at least
one of which must belong to $\frG_{m-1}'$.  We obtain
\ba
\int_0^\beta \dd\tau &\sum_{B, \bar B \subset \caQ} (\|K_B\| + \|K_B''\|)
\e{2\nu
\ell^\nu (\log M + \tau) |\bar B|} \int_0^\beta \dd\tau' \e{-\Delta_0
\tau'} \int_{\frG_{m-1}'} \dd\caG \tilde
z(\caG) \nn\\
&\hspace{4.5cm} \sum_{k\geq0} \frac1{k!} \Bigl( 2\sum_{y \in \bar B}
\int_0^\beta \dd\tau \e{-\Delta_0 \tau} \int_{\frG_{m-1,0}} \dd\caG \tilde
z(\caG)
\Bigr)^k \nn \\
&\leq \beta |\caQ| \frac{(\|\bsK\|_c + \|\bsK''\|_c)^2}{\Delta_0}.
\end{align}
Plugging these estimates in \eqref{boundzA}, one easily gets
\be
z(\caA) \e{\tau|A|} \leq \e{-\beta e_0^\bsmu |A|} \e{3|A|}.
\end{equation}
Exponential decay of the weights of the contours is now clear.

The bound on the derivative can be proven in the same way. Looking at \eqref{defweight}, we
see that the integrand gets a factor bounded by $\beta |A| \sup_{x,\bsmu,\omega,j}
|\frac\partial{\partial\bsmu_j} \Phi_x(\omega_{U(x)})|$.

\subsection{Other properties of the weights}
\label{secrest}

The weight of the contours can be viewed as a series in powers of
$\{K_B\}$, $\{K_B'\}$, $\{K_B''\}$.  Since it is absolutely convergent
uniformly in $\bsK$, $\bsK'$, $\bsK''$ (provided they be small enough), we
have by the dominated convergence theorem
\be
\lim_{\bsK, \bsK',\bsK'' \to 0} z(\caA) = 0.
\end{equation}

Analyticity of $z(\caA)$ as function of $\bsmu$ and $\beta$ is clear, as
well as a function of $\eta$ if we add a new perturbation $\eta\bsL$, in a
neighborhood of 0 that depends on $\beta \|\bsL\|$.  Periodicity is also
obvious.

\vspace{4mm}
{\it Acknowledgments:} It is a pleasure to thank Roberto Fern\'andez, Roman Koteck\'y, and
Charles-\'Edouard Pfister for several discussions, and the referee for useful comments.

\end{document}